\documentclass[entropy,article,accept,moreauthors,pdftex,10pt,a4paper]{mdpi}
\firstpage{1}
\articlenumber{x}
\doinum{10.3390/------}
\pubvolume{19}
\pubyear{2017}
\copyrightyear{2017}
\externaleditor{}
\history{Received: 27 June 2017 ; Accepted: 8 August 2017 ; Published: 21 August 2017}

\usepackage{dcolumn}	
\usepackage{bm}			
\usepackage{amssymb}    
\usepackage{nicefrac}	
\usepackage{verbatim}   
\usepackage[utf8]{inputenc}
\usepackage{soul}

\setitemize{parsep=6pt,itemsep=0pt,leftmargin=*,labelsep=5.5mm}
\setenumerate{parsep=6pt,itemsep=0pt,leftmargin=*,labelsep=5.5mm}
\setlist[description]{itemsep=0mm}

\newcommand{\avg}[1]{\left< #1 \right>}

\newcommand{\mb}{\mathbf}

\usepackage{microtype}
\usepackage{soul}

\Title{Minimum and Maximum Entropy Distributions for Binary Systems with Known Means and Pairwise~Correlations}

\Author{Badr F. Albanna $^{1,2,3}$, Christopher Hillar $^{3,4}$,
Jascha Sohl-Dickstein $^{3,5,6,7}$ and \mbox{Michael R. DeWeese $^{2,3,5,}$*}}

\AuthorNames{Badr F. Albanna, Christopher Hillar, Jascha Sohl-Dickstein and Michael R. DeWeese}

\address{%
$^{1}$ \quad Department of Natural Sciences, Fordham University, New York, NY 10023, USA; balbanna@fordham.edu \\
$^{2}$ \quad Department of Physics, University of California, Berkeley, CA 94720, USA \\
$^{3}$ \quad Redwood Center for Theoretical Neuroscience, University of California, Berkeley, CA 94720, USA; chillar@msri.edu (C.H.); jaschasd@google.com (J.S.-D.) \\
$^{4}$ \quad Mathematical Sciences Research Institute, Berkeley, CA 94720, USA \\
$^{5}$ \quad Helen Wills Neuroscience Institute, University of California, Berkeley, CA 94720, USA\\
$^{6}$ \quad Biophysics Graduate Group, University of California, Berkeley, CA 94720, USA \\
$^{7}$ \quad Google Brain, Google, Mountain View, CA 94043, USA }

\corres{Correspondence: deweese@berkeley.edu}

\abstract{Maximum entropy models are increasingly being used to describe the collective activity of neural populations with measured mean neural activities and pairwise correlations, but the full space of probability distributions consistent with these constraints has not been explored.  We provide upper and lower bounds on the entropy for the {\em minimum} entropy distribution over arbitrarily large collections of binary units with any fixed set of mean values and pairwise correlations. We also construct specific low-entropy distributions for several relevant cases.  Surprisingly, the minimum entropy solution has entropy scaling logarithmically with system size for any set of first- and second-order statistics consistent with arbitrarily large systems. We further demonstrate that some sets of these low-order statistics can only be realized by small systems. Our results show how only small amounts of randomness are needed to mimic low-order statistical properties of highly entropic distributions, and we discuss some applications for engineered and biological information transmission systems}


\begin{document}

\section{Introduction}

Maximum entropy models are central to the study of physical systems in thermal equilibrium~\cite{Pathria:1972p5861},
and they have recently been found to model protein folding~\cite{Russ_2005,Socolich_2005}, antibody diversity~\cite{mora_2010}, and~neural population activity~\cite{Schneidman_Nature_2006,Shlens_JN_2006,Tkacik_2006,tang_2008,Shlens:2009p4887} quite well (see \cite{ganmor_2011} for a~different finding).  In part due to this success, these~types of models have also been used to infer functional connectivity in complex neural circuits~\mbox{\cite{yu_2008, koster2013,hamilton_2013}} and to model collective phenomena of systems of organisms, such as flock behavior~\cite{bialek_flock_of_birds_2011}.

This broad application of maximum entropy models is perhaps surprising since the usual physical arguments involving ergodicity or equality among energetically accessible states are not obviously applicable for such systems, though maximum entropy models have been justified in terms of imposing no structure beyond what is explicitly measured~\cite{jaynes_1957_I,Schneidman_Nature_2006}.
With this approach, {the choice of measured constraints} fully specifies the corresponding maximum entropy model.
More generally, choosing a~set of constraints restricts the set of consistent probability distributions---the maximum entropy solution being only one of all possible consistent probability distributions.
If the space of distributions were sufficiently constrained by observations so that only a~small number of very similar models were consistent with the data, then agreement between the maximum entropy model and the data would be an unavoidable consequence of the constraints rather than a~consequence of the unique suitability of the maximum entropy model for the dataset in question.

In the field of systems neuroscience, understanding the range of allowed entropies for given constraints is an area of active interest.  For example, there has been controversy~\cite{Schneidman_Nature_2006,bethge_2008,roudi_nirenberg_latham_2009,nirenberg_victor_2007,azhar_2010} over the notion that small pairwise correlations can conspire to constrain the behavior of large neural ensembles, which has led to speculation about the possibility that groups of $\sim$200 neurons or more might employ a~form of error correction when representing sensory stimuli~\cite{Schneidman_Nature_2006}.
Recent work~\cite{tkacik_searching_2014} has extended the number of simultaneously recorded neurons modeled using pairwise {statistics} up to 120 neurons, pressing closer to this predicted error correcting limit.
This controversy is in part a~reflection of the fact that  pairwise models do not always allow accurate extrapolation from small populations to large ensembles~\cite{bethge_2008,roudi_nirenberg_latham_2009}, pointing to the need for exact solutions in the important case of large neural systems.
Another recent paper~\cite{Macke2011} has also examined specific classes of biologically-plausible neural models whose entropy grows linearly with system size. Intriguingly, these authors point out that
entropy can be subextensive, at least for one special distribution that is not well matched to most neural data (coincidentally, that special case was originally studied nearly 150 years ago~\cite{Sylvester1867}
when entropy was a~new concept, though not in the present context).
Understanding the range of possible scaling properties of the entropy in a~more general setting is of particular importance to neuroscience because of its interpretation as a~measure of the amount of information communicable by a~neural system to groups of downstream neurons.

Previous authors have studied the large scale behavior of these systems with maximum entropy models expanded to second-~\cite{Schneidman_Nature_2006}, third-~\cite{roudi_nirenberg_latham_2009}, and fourth-order~\cite{azhar_2010}.
Here we use non-perturbative methods to derive rigorous upper and lower bounds on the entropy of the \emph{minimum} entropy distribution for {\em any} fixed sets of means and pairwise correlations possible for arbitrarily large systems
(Equations~\eqref{eqn:min_summary}--\eqref{eqn:exch_summary}; Figures {\ref{fig:SvN} and \ref{fig:Svnu}}).
We also derive lower bounds on the {\em maximum} entropy distribution (Equation~\eqref{eqn:max_summary}) and construct explicit low and high entropy models (Equations~\eqref{H_for_con} and \eqref{H_for_con2}) for a~broad array of cases including the full range of possible uniform first- and second-order constraints realizable for large systems.
\
Interestingly, we find that entropy differences between models with the same first- and second-order statistics can be nearly as large as is possible between any two arbitrary distributions over the same number of binary variables, provided that the solutions do not run up against the boundary of the space of allowed constraint values (see {Section~{\ref{sec:LimSysGro}}} and Figure~\ref{fig:crash} for a~simple illustration of this phenomenon).
This boundary is structured in such a~way that some ranges of values for low order statistics are only satisfiable by systems below some critical size.
Thus, for cases away from the boundary, entropy is only weakly constrained by these statistics, and the success of maximum entropy models in biology~\cite{Russ_2005,Socolich_2005,mora_2010,Schneidman_Nature_2006,Shlens_JN_2006,Tkacik_2006,tang_2008,yu_2008,Shlens:2009p4887,bialek_flock_of_birds_2011}, when it occurs for large enough systems~\cite{roudi_nirenberg_latham_2009}, can represent a~real triumph of the maximum entropy approach.

Our results also have relevance for engineered information transmission systems.
We show that empirically measured first-, second-, and even third-order statistics are essentially inconsequential for testing coding optimality in a~broad class of such systems, whereas the existence of other statistical properties, such as finite exchangeability~\cite{Diaconis_1977}, do guarantee information transmission near channel capacity~\cite{shannon1948, cover_thomas}, the maximum possible information rate given the properties of the information channel.
A better understanding of minimum entropy distributions subject to constraints is also important for minimal state space realization \cite{BDe2000331, Shalizi2001}---a form of optimal model selection based on an~interpretation of Occam's Razor complementary to that of Jaynes~\cite{jaynes_1957_I}. Intuitively, maximum entropy models impose no structure beyond that needed to fit the measured properties of a~system, whereas minimum entropy models require the fewest ``moving parts'' in order to fit the data.  In~addition, our~results have implications for computer science as algorithms for generating binary random variables with constrained statistics and low entropy have found many applications ({e.g.}, \cite{carter1979universal, sipser1983complexity, stockmeyer1983complexity, chor1985bit, karp1985fast, luby86, alon1986fast, alexi1988rsa, chor1989power, berger1991simulating, schulman1992sample, luby1993removing, motwani1994probabilistic, koller1993constructing, karloff1994construction,castellana_inverse_2014}).

\subsection{Problem Setup}

To make these ideas concrete,
consider an abstract description of a~neural ensemble consisting of $N$ spiking neurons.
In any given time bin, each neuron $i$ has binary state $s_i$ denoting whether it is currently firing an action potential ($s_i = 1$) or not ($s_i = 0$).  The state of the full network is represented by $\vec{s} = (s_1,\ldots,s_N) \in \{0,1\}^N$.  Let $p(\vec{s})$ be the probability of state $\vec{s}$ so that the distribution over all $2^N$ states of the system is represented by $\textbf{p} \in [0, 1]^{2^N}$, $\sum_{\vec{s}} p(\vec{s}) = 1$. {Although we will use this neural framework throughout the paper, note that all of our results will hold for any type of system consisting of binary variables.}

In neural studies using maximum entropy models, electrophysiologists typically measure the time-averaged firing rates $\mu_i = \avg{s_i}$ and pairwise event rates $\nu_{ij} = \avg{s_i s_j}$ and fit the maximum entropy model consistent with these constraints,  yielding a~Boltzmann distribution for an Ising spin glass~\cite{hertz_spin_glass}.
This ``inverse'' problem of inferring the interaction and magnetic field terms in an Ising spin glass Hamiltonian that produce the measured means and correlations is nontrivial, but there has been progress~\cite{Tanaka:1998p1984,Hinton2006,Hyvarinen:2007p5984,Broderick:2007p2761,azhar_2010,MPF_PRL,Tkacik_2013}.
The maximum entropy distribution is not the only one consistent with these observed statistics, however.
In fact, there are typically many {such distributions}, and we will refer to the complete set of these as the \textit{solution space} for a~given set of constraints.
Little is known about the minimum entropy permitted for a~particular solution space.

Our question is: Given a~set of observed mean firing rates and pairwise correlations between neurons, what are the possible entropies for the system?  We will denote the maximum (minimum) entropy compatible with a~given set of imposed correlations up to order $n$ by $S_n$ ($\tilde{S}_n$).
The maximum entropy framework~\cite{Schneidman_Nature_2006} provides a~hierarchical representation of neural activity: as increasingly higher order correlations are measured, the corresponding model entropy $S_n$ is reduced (or remains the same) until it reaches a~lower limit.  Here we introduce a~complementary, minimum entropy framework: as~higher order correlations are specified, the corresponding model entropy $\tilde{S}_n$ is increased (or unchanged) until all correlations are known.
The range of possible entropies for any given set of constraints is the gap ($S_n - \tilde{S}_n$) between these two model entropies, and our primary concern is whether this gap is greatly reduced for a~given set of observed first- or second-order statistics. We find that, for many cases, the gap grows linearly with the system size $N$, up to a~logarithmic correction.

\section{Results}

We prove the following bounds on the minimum and maximum entropies for fixed sets of values of first and second order statistics, {$\{\mu_i\} = \{\avg{s_i}\}$}  and {$\{\nu_{ij}\} = \{\avg{s_i s_j}\}$}, respectively.  All entropies are given in bits.

For the \emph{minimum entropy}:
\begin{equation}
    \log_2 \left( \frac{N}{1 + (N-1) \bar\alpha } \right) \leq \tilde{S}_2 \leq \log_2 \left( 1+\frac{N(N+1)}{2} \right),
    \label{eqn:min_summary}
\end{equation}
where $\bar\alpha$ is the average of $\alpha_{ij} = (4\nu_{ij} - 2\mu_i - 2\mu_j + 1)^2$ over all $i,j \in \{1,\dots,N\}$, $i \neq j$.
Perhaps surprisingly, the scaling behavior of the minimum entropy does not depend on the details of the sets of constraint values --- for large systems the entropy floor does not contain tall peaks or deep valleys as one varies $\{\mu_i\}$, $\{\nu_{ij}\}$, or $N$.

We emphasize that the bounds in Equation~(\ref{eqn:min_summary}) are valid for arbitrary sets of mean firing rates and pairwise correlations, but
we will often focus on the special class of distributions with {\em uniform~constraints}:
\begin{align}
\label{eq:sym_mu}
  \mu_i &= \mu, \quad \text{for all } i = 1,\ldots,N \\
\label{eq:sym_nu}
  \nu_{ij} &= \nu, \quad \text{for all } i \neq j.
\end{align}

The allowed values for $\mu$ and $\nu$ that can be achieved for arbitrarily large systems (see Appendix~\ref{sec:range_of_mu_nu})~are
\begin{equation}
\label{eq:allowed_mu_nu}
\mu^2 \leq \nu \leq \mu.
\end{equation}

For uniform constraints, Equation~\eqref{eqn:min_summary} reduces to
\begin{equation}
\log_2 \left( \frac{N}{1+ (N-1) \alpha} \right) \leq \tilde{S}_2 \leq \log_2 \left( 1+\frac{N(N+1)}{2} \right),
\end{equation}
where $\alpha(\mu,\nu) = (4(\nu-\mu) + 1)^2$. In most cases, the lower bound in Equation~\eqref{eqn:min_summary} asymptotes to a~constant, but in the special case for which $\mu$ and $\nu$ have values consistent with the global maximum entropy solution ($\mu = \nicefrac{1}{2}$ and $\nu = \nicefrac{1}{4}$), we can give the higher
bound:
\begin{equation}
	\log_2 (N) \leq \tilde{S}_2 \leq \log_2 (N) + 2.
\label{eqn:min_summary_spec}
\end{equation}

For the \emph{maximum entropy}, it is well known that:
\begin{equation}
S_2 \leq N,
\end{equation}
which is valid for any set of values for $\{\mu_i\}$ and $\{\nu_{ij}\}$.
Additionally, for any set of uniform constraints that can be achieved by arbitrarily large systems (Equation~(\ref{eq:allowed_mu_nu})) other than $\nu = \mu$, which corresponds to the case in which all $N$ neurons are perfectly correlated,
we can derive upper and lower bounds on the maximum entropy that each scale linearly with the system size:
\begin{equation}
	w + x N  \leq S_2 \leq N,
	\label{eqn:max_summary}
  \end{equation}
where $w = w(\mu,\nu) \geq 0$ and $x = x(\mu,\nu) > 0$ are independent of $N$.

{An important class of probability distributions are the \emph{exchangeable distributions}~{\cite{Diaconis_1977}}, which are distributions over multiple variables that are symmetric under any permutation of those variables. For binary systems, exchangeable distributions have the property that the probability of a~sequence of ones and zeros is only a~function of the number of ones in the binary string.}
We have constructed a~family of exchangeable distributions, with entropy $\tilde{S}^{exch}_2$, that we conjecture to be minimum entropy exchangeable solutions.  The entropy of our exchangeable constructions scale linearly with $N$:
\begin{equation}
C_1 N - \mathcal{O}(\log_2 N) \leq \tilde{S}^{exch}_2 \leq C_2 N,
\label{eqn:exch_summary}
\end{equation}
where $C_1 = C_1(\mu,\nu)$ and $C_2 = C_2(\mu,\nu)$ do not depend on $N$.
We have computationally confirmed that this is indeed a~minimum entropy exchangeable solution for $N \leq 200$.

\begin{figure}[H]
\centering
\includegraphics[width=0.55\linewidth]{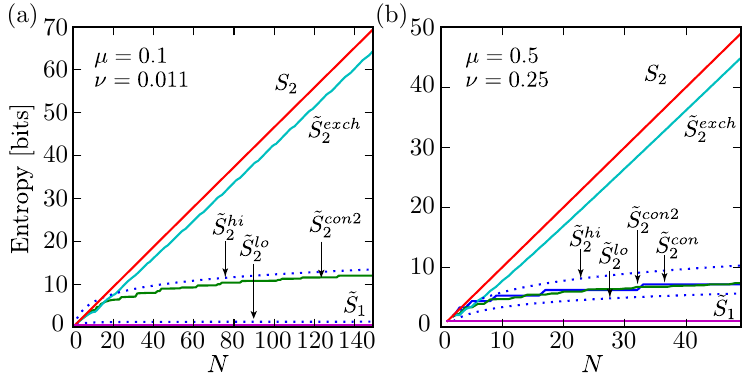}
\caption{
Minimum and maximum entropy for fixed uniform constraints as a~function of $N$.
The~minimum entropy grows no faster than logarithmically with the system size $N$ for any mean activity level $\mu$ and pairwise correlation strength $\nu$.
(\textbf{a}) In a~parameter regime relevant for neural population activity in the retina~\cite{Schneidman_Nature_2006,Shlens_JN_2006} ($\mu = 0.1$, $\nu = 0.011$), we can construct an explicit low entropy solution ($\tilde{S}_2^{con2}$) that grows logarithmically with $N$,  unlike the linear behavior of the maximum entropy solution ($S_2$).  Note that the linear behavior of the maximum entropy solution is only possible because these parameter values remain within the boundary of allowed $\mu$ and $\nu$ values (See Appendix~\ref{sec:max_entropy}); (\textbf{b})~Even~for mean activities and pairwise correlations matched to the global maximum entropy solution ($S_2$; $\mu = \nicefrac{1}{2}$, $\nu = \nicefrac{1}{4}$), we can construct explicit low entropy solutions ($\tilde{S}_2^{con}$ and $\tilde{S}_2^{con2}$) and a~lower bound ($\tilde{S}_2^{lo}$) on the entropy that each grow logarithmically with $N$, in contrast to the linear behavior of the maximum entropy solution ($S_2$)
and the finitely exchangeable minimum entropy solution ($\tilde{S}_2^{exch}$). $\tilde{S}_1$ is the minimum entropy distribution that is consistent with the mean firing rates.  It remains constant as a~function of $N$.}
\label{fig:SvN}
\end{figure}

\begin{figure}[H]
\centering
\includegraphics[width=0.5\linewidth]{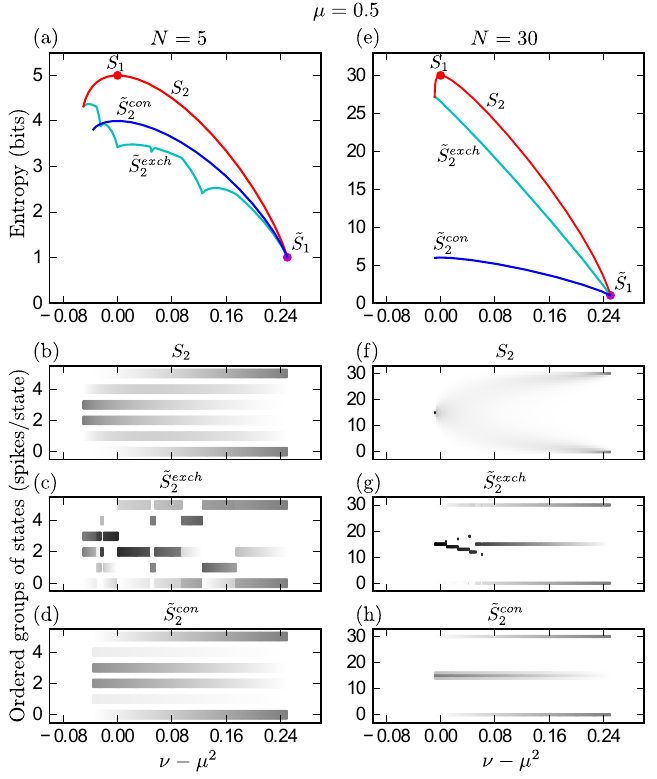}
\caption{
Minimum and maximum entropy models for uniform constraints. (\textbf{a}) Entropy as a~function of the strength of pairwise correlations for the maximum entropy model ($S_2$), finitely exchangeable minimum entropy model ($\tilde{S}^{exch}_2$), and a~constructed low entropy solution ($\tilde{S}^{con}_2$), all corresponding to $\mu = \nicefrac{1}{2}$ and $N=5$.
Filled circles indicate the global minimum $\tilde{S}_1$ and maximum $S_1$ for $\mu = \nicefrac{1}{2}$;
(\textbf{b}--\textbf{d})~Support for $S_2$ (b), $\tilde{S}^{exch}_2$ ({c}), and $\tilde{S}^{con}_2$ ({d}) corresponding to the three curves in panel ({a}).
States are grouped by the number of active units; darker regions indicate higher total probability for each group of states; (\textbf{e}--\textbf{h}) Same as for panels ({a}) through ({d}), but with $N = 30$. Note that, with rising $N$,
the cusps in the $\tilde{S}_2^{exch}$ curve become much less pronounced.
}
\label{fig:Svnu}
\end{figure}

Figure~\ref{fig:SvN} illustrates the scaling behavior of these various bounds for uniform constraints in two parameter regimes of interest.
Figure~{\ref{fig:Svnu}} shows how the entropy depends on the level of correlation ($\nu$) for the maximum entropy solution ($S_2$), the minimum entropy exchangeable solution($\tilde{S}^{exch}_2$), and a~low entropy solution ($\tilde{S}^{con}_2$), for a~particular value of mean activity ($\mu = \nicefrac{1}{2}$) at each of two system sizes, $N = 5$ and $N = 30$.
\vspace{-18pt}

\subsection{Limits on System Growth}
\label{sec:LimSysGro}

Physicists are often faced with the problem of having to determine some set of experimental predictions ({e.g.}, mean values and pairwise correlations of spins in a~magnet) for some model system defined by a~given Hamiltonian, which specifies the energies associated with any state of the system.
Typically, one is interested in the limiting behavior as the system size tends to infinity. In this canonical situation, no matter how the various terms in the Hamiltonian are defined as the system grows, the~existence of a~well-defined Hamiltonian guarantees that there exists a~(maximum entropy) solution for the probability distribution for any system size.

However, here we are studying the {\em inverse} problem of deducing the underlying model based on measured low-order statistics~\cite{Tanaka:1998p1984,Hinton2006,Hyvarinen:2007p5984,Broderick:2007p2761,azhar_2010,MPF_PRL}.
In particular, we are interested in the minimum and maximum entropy models consistent with a~given set of specified means and pairwise correlations. Clearly, both of these types of (potentially degenerate) models must exist whenever there exists at least one distribution consistent with the specified statistics, but as we now show, some sets of constraints can only be realized for small systems.

\subsubsection{A Simple Example}
\begin{figure}[H]
\centering
\includegraphics[width=0.53\linewidth]{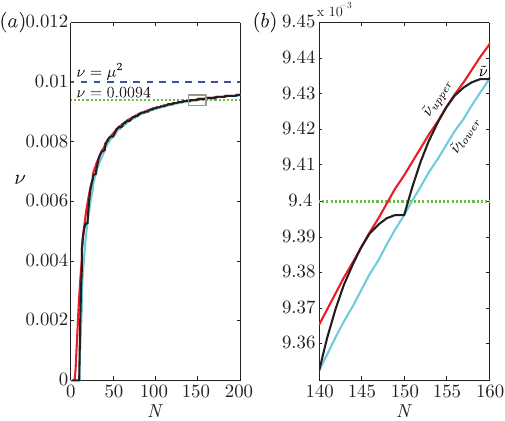}
\caption{
An example of uniform, low-level statistics that can be realized by small groups of neurons but not by any system greater than some critical size. (\textbf{a}) Upper (red curve, $\tilde{\nu}_{upper}$) and lower (cyan curve, $\tilde{\nu}_{lower}$) bounds on the minimum value (black curve, $\tilde{\nu}$) for the pairwise correlation $\nu$ shared by all pairs of neurons are plotted as a~function of system size $N$ assuming that every neuron has mean activity $\mu = 0.1$. Note that all three curves asymptote to $\nu = \mu^2 = 0.01$ as $N \rightarrow \infty$ (dashed blue line); (\textbf{b}) Enlarged portion of (\textbf{a}) outlined in grey reveals that groups of $N \le 150$ neurons can exhibit uniform constraints $\mu = 0.1$ and $\nu = 0.0094$ (green dotted line), but this is not possible for any larger group.
}
\label{fig:crash}
\end{figure}

To illustrate this point, consider the following example, which is arguably the simplest case exhibiting a~cap on system size.
At least one system consisting of $N$ neurons can be constructed to satisfy the uniform set of constraints: $\mu = 0.1$ and $\nu = 0.0094 < \mu^2 = 0.01$, provided that $2 \le N \le 150$, but no solution is possible for $N > 150$.
To prove this, we first observe that any set of uniform constraints that admits at least one solution must admit at least one exchangeable solution (see~Appendices~\ref{sec:range_of_mu_nu} and \ref{sec:exch_entropy}). Armed with this fact, we can derive closed-form solutions for upper and lower bounds on the minimum value for $\nu$ consistent with $N$ and $\mu$ (Appendix Equation~(\ref{ineq:nu_min_bounds})), as~well as the actual minimum value of $\nu$ for any given value of $N$, as depicted in Figure~\ref{fig:crash}.

Thus, this system defined by its low-level statistics cannot be grown indefinitely, not because the entropy decreases beyond some system size, which is impossible ({e.g.}, Theorem 2.2.1 of~\cite{cover_thomas}), but~rather, because no solution of any kind is possible for these particular constraints for ensembles of more than 150 neurons.  We believe that this relatively straightforward example captures the essence of the more subtle phenomenon described by {Schneidman} and colleagues~\cite{Schneidman_Nature_2006}, who pointed out
that a~naive extrapolation of the maximum entropy model fit to their retinal data to larger system sizes results in a~conundrum at around $N \approx 200$ neurons.

For our simple example, it was straightforward to decide what rules to follow when growing the system---both the means and pairwise correlations were fixed and perfectly uniform across the network at every stage.  In practice, real neural activities and most other types of data exhibit some variation in their mean activities and correlations across the measured population, so in order to extrapolate to larger systems, one must decide how to model the distribution of statistical values involving the added neurons.

Fortunately, despite this complication, we have been able to derive upper and lower bounds on the minimum entropy for arbitrarily large $N$. In Appendix~\ref{sec:max_entropy}, we derive a~lower bound on the maximum entropy for the special case of uniform constraints achievable for arbitrarily large systems, but such a~bound for the more general case would depend on the details of how the statistics of the system change with $N$.

Finally, we mention that this toy model can be thought of as an example of a~``frustrated'' system~\cite{Toulouse_1980}, in that traveling in a~closed loop of an odd number of neurons involves an odd number of anticorrelated pairs. By ``anticorrelated'', we mean that the probability of simultaneous firing of a~pair of neurons is less than chance, $\nu < \mu^2$, but note that $\nu$ is always non-negative due to our convention of labeling active and inactive units with zeros and ones, respectively. However, frustrated systems are not typically defined by their correlational structure, but rather by their Hamiltonians, and there are often many Hamiltonians that are consistent with a~given set of observed sets of constraints, so there does not seem to be a~one-to-one relationship between these two notions of frustration.
In the more general setting of nonuniform constraints, the relationship between frustrated systems as defined by their correlational structure and those that cannot be grown arbitrarily large is much more complex.

\subsection{Bounds on Minimum Entropy}

Entropy is a~strictly concave function of the probabilities and therefore has a~unique maximum that can be identified using standard methods \cite{Boyd:2004:CO:993483}, at least for systems that possess the right symmetries or that are sufficiently small.
In {Section~{\ref{sec:maxent}}}, we will show that the maximum entropy $S_2$ for many systems with specified means and pairwise correlations scales linearly with $N$ (Equation~\eqref{eqn:max_summary}, Figure~\ref{fig:SvN}).
We obtain bounds on minimum entropy by exploiting the geometry of the entropy function.

\subsubsection{Upper Bound on the Minimum Entropy}

We prove below that the minimum entropy distribution exists at a~vertex of the allowed space of probabilities, where most states have probability zero \cite{Rosen}.
Our challenge then is to determine in which vertex a~minimum resides.
The entropy function is nonlinear, precluding obvious approaches from linear programming, and the dimensionality of the probability space grows exponentially with $N$, making exhaustive search and gradient descent techniques intractable for $N \gtrsim 5$.
Fortunately, we can compute a~lower (upper) bound $\tilde{S}^{lo}_2$ ($\tilde{S}^{hi}_2$) on the entropy of the minimum entropy solution for all $N$ (Figure~\ref{fig:SvN}), and we have constructed two families of explicit solutions with low entropies ($\tilde{S}^{con}_2$ and $\tilde{S}^{con2}_2$; Figures~\ref{fig:SvN} and \ref{fig:Svnu}) for a~broad parameter regime covering all allowed values for $\mu$ and $\nu$ in the case of uniform constraints that can be achieved by arbitrarily large systems (see Equation~\eqref{eqn:range_of_nu}, Figure~\ref{fig:allowed} in the Appendix).

Our goal is to minimize the entropy $S$ as a~function of the probabilities $p_i$ where
\begin{equation}
S(\mathbf{p}) = -\sum_{i=1}^{n_s} p_i \log_2 p_i,
\end{equation}
$n_s$ is the number of states, the $p_i$ satisfy a~set of $n_c$ independent linear constraints, and $p_i \geq 0$ for all $i$.
For the main problem we consider, $n_s = 2^N$.  The number of constraints---normalization, mean firing rates, and pairwise correlations---grows quadratically with $N$:
\begin{align}
n_c  = 1 + \frac{N(N+1)}{2}.
\end{align}

The space of normalized probability distributions $\mathcal{P} = \{p: \sum_{i = 1}^{n_s} {p_i} = 1, \ p_i \geq 0\}$ is the standard simplex in $n_s - 1$ dimensions.
Each additional linear constraint on the probabilities introduces a~hyperplane in this space.  If the
constraints are consistent and independent, then the intersection of these hyperplanes
defines a~$d = n_s - n_c$ affine space, which we call $\mathcal{C}$.
All solutions are constrained to the intersection between $\mathcal{P}$ and $\mathcal{C}$, and this solution space is a~convex polytope of dimension $\le$$d$, which~we refer to as
$\mathcal{R}$.
A point within a~convex polytope can always be expressed as a~linear combination of its vertices; therefore, if $\{\mathbf{v}_i\}$ are the vertices of
$\mathcal{R}$, we may express
\begin{equation}
\mathbf{p} = \sum_i^{n_v} a_i \mathbf{v}_i,
\end{equation}
where $n_v$ is the total number of vertices and $\sum_i^{n_v} a_i = 1$.

Using the concavity of the entropy function, we will now show that the minimum entropy for a~space of probabilities $S$ is attained on one (or possibly more) of the vertices of that space.
Moreover, these vertices correspond to probability distributions with small support---specifically a~support size no greater than $n_c$ (see Appendix~\ref{sec:support}).
This means that the global minimum will occur at the (possibly degenerate) vertex that has the lowest entropy, which we denote as $\mathbf{v}_*$:
\begin{align}
S(\mathbf{p}) & = S\left(\sum_i^{n_v} a_i \mathbf{v}_i\right) \nonumber\\
              & \geq \sum_i^{v_s} a_i S(\mathbf{v}_i) \nonumber\\
              & \geq \left( \sum_i^{v_s} a_i \right) S(\mathbf{v}_*) \nonumber\\
              & = S(\mathbf{v}_*).
\end{align}

It follows that $\tilde{S}_2 = S(\mathbf{v}_*)$.

Moreover, if a~distribution satisfying the constraints exists, then there is one with at most $n_c$ nonzero $p_i$ ({e.g.}, from arguments as in \cite{koller1993constructing}).  Together, these two facts imply that there are minimum entropy distributions with a~maximum of $n_c$ nonzero $p_i$.
This means that even though the state space may grow exponentially with $N$, the support of the minimum entropy solution for fixed means and pairwise correlations will only scale quadratically with $N$.

The maximum entropy possible for a~given support size occurs when the probability distribution is evenly distributed across the support and the entropy is  equal to the logarithm of the number of states. This allows us to give an upper bound on the minimum entropy as
\begin{align}
\tilde{S}_2 \leq \tilde{S}_2^{hi}   & = \log_2(n_c) \nonumber\\
                                    & = \log_2\left(1 + \frac{N(N+1)}{2}\right) \nonumber\\
                                    & \approx 2 \log_2 (N), \quad N \gg 1.
\end{align}

Note that this bound is quite general: as long as the constraints are independent and consistent this result holds regardless of the specific values of  the $\{\mu_i\}$ and $\{\nu_{ij}\}$.

\subsubsection{Lower Bound on the Minimum Entropy}

We can also use the concavity of the entropy function to derive a~lower bound on the entropy $\tilde{S}^{lo}_2$ as in Equation \eqref{eqn:min_summary}:
\begin{equation}
\tilde{S}_2(N,\{\mu_i\},\{\nu_{ij}\}) \geq \tilde{S}^{lo}_2 = 	\log_2 \left( \frac{N^2}{N +\sum_{i \neq j} \alpha_{ij}} \right),
\label{eq:S2_low_bound}
\end{equation}
where $\tilde{S}_2(N,\{\mu_i\},\{\nu_{ij}\})$ is the minimum entropy given a~network of size $N$ with constraint values $\{\mu_i\}$ and $\{\nu_{ij}\}$,  and the sum is taken over all $i,j \in \{1,\dots,N\}$, $i \neq j$ and $\alpha_{ij} = (4\nu_{ij} - 2\mu_i - 2\mu_j + 1)^2$.
Taken together with the upper bound, this fully characterizes the scaling behavior of the entropy floor as a~function of $N$.

To derive this lower bound, note that the concavity of the entropy function allows us to write

\begin{equation}
S(\mathbf{p}) \geq -\log_2 \|{\mathbf{p}}\|_2^2.
\label{eqn:p2_bound}
\end{equation}

Using this relation to find a~lower bound on $S(\mathbf{p})$ requires an upper bound on $\| \mathbf{p} \|_2^2$ provided by the Frobenius norm of the correlation matrix $C \equiv \langle \vec{s} \vec{s}^T \rangle$ (where the states are defined to take vaues in $\{-1,1\}^N$ rather than $\{0,1\}^N$)

\begin{equation}
\| \mathbf{p} \|_2^2 \leq \frac{\| {C} \|^2_F}{N^2}.
\label{eqn:frobenius_bound}
\end{equation}

In this case, $\| {C} \|^2_F$ is a~simple function of the $\{\mu_i\}$, $\{\nu_{ij}\}$:

\begin{equation}
\| {C} \|^2_F = N +\sum_{i \neq j} (4\nu_{ij} - 2\mu_i - 2\mu_j + 1)^2.
\label{eqn:frobenius}
\end{equation}

\textls[-15]{Using Equations~(\ref{eqn:frobenius_bound}) and (\ref{eqn:frobenius}) in Equation~(\ref{eqn:p2_bound})  gives us our result (see Appendix~\ref{sec:lower_bound} for further~details).}

Typically, this bound asymptotes to a~constant as the number of terms included in the sum over $\alpha_{ij}$ scales with the number of pairs ($\mathcal{O}(N^2)$), but in certain conditions this bound can grow with the system size.
For example, in the case of uniform constraints, this reduces to

\begin{equation}
\tilde{S}^{lo}_2 = \log_2 \left( \frac{N}{1+ (N-1) \alpha} \right),
\end{equation}
where $\alpha(\mu,\nu) = (4(\nu-\mu) + 1)^2$.

In the special case

\begin{equation}\label{alpha_zero}
\nu = \mu - \nicefrac{1}{4},
\end{equation}
$\alpha$ vanishes allowing the bound in Equation~(\ref{eq:S2_low_bound}) to scale logarithmically with $N$.

In the large $N$ limit for uniform constraints, we know $\mu \geq \nu \geq \mu^2$ (see Appendix~\ref{sec:range_of_mu_nu}), therefore the only values of $\mu$ and $\nu$ satisfying Equation~(\ref{alpha_zero}) are

\begin{equation}
\mu = \nicefrac{1}{2}, \quad \nu = \mu^2 = \nicefrac{1}{4}.
\label{eq:munu_max_S}
\end{equation}

Although here the lower bound grows logarithmically with $N$, rather than remaining constant, for many large systems this difference is insignificant compared with the linear dependence $S_0 = N$ of the maximum entropy solution  ({i.e.}, $N$ fair i.i.d. Bernoulli random variables).
In other words, the gap between the minimum and maximum possible entropies consistent with the measured mean activities and pairwise correlations grows linearly in these cases (up to a~logarithmic correction), which is as large as the gap for the space of all possible distributions for binary systems of the same size $N$ with the same mean activities but without restriction on the correlations.

\subsection{Bounds on Maximum Entropy}
\label{sec:maxent}

An upper bound on the maximum entropy is well-known and easy to state.
For a~given number of allowed states $n_s$, the maximum possible entropy occurs when the probability distribution is equally distributed over all allowed states ({i.e.}, the microcanonical ensemble), and the entropy is equal to
\begin{align}
S_2 \leq S_2^{hi} 	& = \log_2(n_s) \nonumber\\
					& = N.
\end{align}

Lower bounds on the maximum entropy are more difficult to obtain
(see Appendix~\ref{sec:max_entropy} for further discussion on this point.). However, by restricting ourselves to uniform constraints achievable by arbitrarily large systems (Equations~(\ref{eq:sym_mu}) and (\ref{eq:sym_nu})), we can construct a~distribution that provides a~useful lower bound on the maximum possible entropy consistent with these constraints:
\begin{equation}
  \label{eqn:S_con_3_maintxt}
    S_2 \geq S^{con}_2 = w + xN,
\end{equation}
where
\begin{align}
  \label{eqn:w_def_maintxt}
  w = & -\beta \log_2\beta - (1 - \beta)\log_2 (1 - \beta), \\
  x  = & (1-\beta) \left[-\eta\log_2\eta - (1 - \eta)\log_2(1 - \eta)\right]
\end{align}
and
\begin{align}
    \beta &= \frac{\nu - \mu^2}{1 + \nu - 2\mu},\\
  \label{eqn:con_3_eta_maintxt}
    \eta &= \frac{\mu - \nu}{1 - \mu}.
\end{align}

It is straightforward to verify that $w$ and $x$ are nonnegative constants for all allowed values of $\mu$ and $\nu$ that can be achieved for arbitrarily large systems. Importantly, $x$ is nonzero provided $\nu < \mu$ \mbox{({ i.e.}, $\beta \neq 1 \cap \eta \neq \{0,1\}$)}, so the entropy of the system will grow linearly with $N$ for any set of uniform constraints achievable for arbitrarily large systems except for the case in which all neurons are perfectly~correlated.

\subsection{Low-Entropy Solutions}

In addition to the bounds we derived for the minimum entropy, we can also construct probability distributions  between these minimum entropy boundaries for distributions with uniform constraints.
These solutions provide concrete examples of distributions
that achieve the scaling behavior of the bounds we have derived for the minimum entropy.
We include these so that the reader may gain a~better intuition for what low entropy models look like in practice.
We remark that these models are not intended as an improvement over maximum entropy models for any particular biological system.
This would be an interesting direction for future work.
Nonetheless, they are of practical importance to other fields as we discuss further in {Section~{\ref{sec:implications}}}.

Each of our low entropy constructions, $\tilde{S}^{con}_2$ and  $\tilde{S}^{con2}_2$, has an entropy that grows logarithmically with $N$ (see Appendices~\ref{sec:prime_construction}--\ref{sec:validity}, Equations \eqref{eqn:min_summary}--\eqref{eqn:min_summary_spec}):
\begin{eqnarray}
\label{H_for_con}
\tilde{S}^{con}_2 &=& \left\lceil \log_2(N) + 1\right\rceil \nonumber \\
&\leq& \log_2(N) + 2,\\
\label{H_for_con2}
\tilde{S}^{con2}_2 &\leq&  \log_2 \left(\left\lceil N\right\rceil_p (\lceil N \rceil_p - 1)\right) - 1 + \log_2(3) \nonumber\\
&\leq& \log_2\left(N(2N - 1)\right) + \log_2(3) ,
\end{eqnarray}
where $\lceil . \rceil$ is the ceiling function and $\lceil . \rceil_p$ represents the smallest prime at least as large as its argument.
Thus, there is always a~solution whose entropy grows no faster than logarithmically with the size of the system, for any observed levels of mean activity and pairwise correlation.

As illustrated in Figure~\ref{fig:SvN}a, for large binary systems with uniform first- and second-order statistics matched to typical values of many neural populations, which have low firing rates and correlations slightly above chance~(\cite{Schneidman_Nature_2006,Shlens_JN_2006,Tkacik_2006,tang_2008,bethge_2008,yu_2008,Shlens:2009p4887}; $\mu = 0.1$, $\nu = 0.011$), the range of possible entropies grows almost linearly with $N$, despite the highly symmetric constraints imposed (Equations~(\ref{eq:sym_mu}) and (\ref{eq:sym_nu})).

Consider the special case of first- and second-order constraints (Equation~(\ref{eq:munu_max_S})) that correspond to the unconstrained global maximum entropy distribution.
For these highly symmetric constraints, both~our upper and lower bounds on the minimum entropy  grow logarithmically with $N$, rather than just the upper bound as we found for the neural regime ({Equation~(\ref{eqn:min_summary_spec})}; Figure~\ref{fig:SvN}a). In fact, one can construct~\mbox{\cite{Macke2011,Sylvester1867}} an~explicit solution (Equation (\ref{H_for_con}); Figures~\ref{fig:SvN}b and \ref{fig:Svnu}a,d,e,h) that matches the mean, pairwise correlations, and triplet-wise correlations of the global maximum entropy solution whose entropy $\tilde{S}^{con}_2$ is never more than two bits above our lower bound (Equation~(\ref{eq:S2_low_bound})) for all $N$.  Clearly then, these~constraints alone do not guarantee a~level of independence of the neural activities commensurate with the maximum entropy distribution.   By varying the relative probabilities of states in this explicit construction we can make it satisfy a~much wider range of $\mu$ and $\nu$ values than previously considered, covering most of the allowed region (see Appendices~\ref{sec:comm_construction} and \ref{sec:validity}) while still remaining a~distribution whose entropy grows logarithmically with $N$.

\subsection{Minimum Entropy for Exchangeable Distributions}

We consider the exchangeable class of distributions as an example of distributions whose entropy must scale linearly with the size of the system unlike the global entropy minimum which we have shown scales  logarithmically.
If one has a~principled reason to believe some system should be described by an exchangeable distribution, the constraints themselves are sufficient to drastically narrow the allowed range of entropies although the gap between the exchangeable minimum and the maximum will still scale linearly with the size of the system except in special cases.
This result is perhaps unsurprising as the restriction to exchangeable distributions is equivalent to imposing a~set of additional constraints ({e.g.}, $p(100)=p(010)=p(001)$, for $N = 3$) that is exponential in the size of the~system.

While a~direct computational solution to the general problem of finding the minimum entropy solution becomes intractable for $N \gtrsim 5$, the situation for the exchangeable case is considerably different.
In this case, the high level of symmetry imposed means that there are only $n_s = N+1$ states (one for each number of active neurons) and $n_c=3$ constraints (one for normalization, mean, and pairwise firing).
This makes the problem of searching for the minimum entropy solution at each vertex of the space computationally tractable up into the hundreds of neurons.

Whereas the global lower bound scales logarithmically, our computation illustrates that the exchangeable case scales with $N$ as seen in Figure~\ref{fig:SvN}.
The large gap between $\tilde{S}^{exch}_2$ and $\tilde{S}_2$ demonstrates that a~distribution can dramatically reduce its entropy if it is allowed to violate sufficiently strong symmetries present in the constraints.
This is reminiscent of other examples of symmetry-breaking in physics for which a~system finds an equilibrium that breaks symmetries present in the physical laws.
However, here the situation can be seen as reversed:
Observed first and second order statistics satisfy a~symmetry that is not present in the underlying model.

\subsection{Implications for Communication and Computer Science}
\label{sec:implications}

These results are not only important for understanding the validity of maximum entropy models in neuroscience, but they also have consequences in other fields that rely on information entropy.
We now examine consequences of our results for engineered communication systems. Specifically, consider a~device such as a~digital camera that exploits compressed sensing~\cite{Candes2005, Donoho2006} to reduce the dimensionality of its image representations.
A compressed sensing scheme might involve taking inner products between the vector of raw pixel values and a~set of random vectors, followed by a~digitizing step to  output $N$-bit strings.  Theorems exist for expected information rates of compressed sensing systems, but we are unaware of any that do not depend on some knowledge about the input signal, such as its sparse structure~\cite{donoho_Stanford_tech_rep_2004,Sarvotham_2006}.
Without such knowledge,
it would be desirable to know which empirically measured output statistics could determine whether such a~camera is utilizing as much of the $N$ bits of channel capacity as possible for each photograph.

As we have shown, even if the mean of each bit is $\mu = \nicefrac{1}{2}$, and the second- and third-order correlations are at chance level ($\nu = \nicefrac{1}{4}$; $\avg{s_i s_j s_k} = \nicefrac{1}{8}$, for all sets of distinct $\{i,j,k\}$), consistent with the maximum entropy distribution, it is possible that the Shannon mutual information shared by the original pixel values and the compressed signal is only on the order of $\log_2(N)$ bits, well below the channel capacity ($N$ bits) of this (noiseless) output stream. We emphasize that, in such a~system, the transmitted information is limited not by corruption due to noise,
which can be neglected for many applications involving digital electronic devices,
but instead by the nature of the second- and higher-order correlations in the output.

Thus, measuring pairwise or even triplet-wise correlations between all bit pairs and triplets is insufficient to provide a~useful floor on the information rate, no matter what values are empirically observed.  However, knowing the extent to which other statistical properties are obeyed can yield strong guarantees of system performance. In particular, exchangeability is one such constraint. Figure~\ref{fig:SvN} illustrates the near linear behavior of the lower bound on information ($\tilde{S}_2^{exch}$) for distributions obeying exchangeability, in both the neural regime (cyan curve, panel (a)) and the regime relevant for our engineering example (cyan curve, panel (b)).
We find experimentally that any exchangeable distribution has as much entropy as the maximum entropy solution, up to terms of order $\log_2(N)$ (see~Appendicies).

This result has potential applications in the field of symbolic dynamics and computational mechanics, which study the consequences of viewing a~complex system through a~finite state measuring device~\cite{Hao1989, Shalizi2001}.
If we view each of the various models presented here as a~time series of binary measurements from a~system, our results indicate that bitstreams with identical mean and pairwise statistics can have profoundly different scaling as a~function of the number of measurements ($N$), indicating radically different complexity.
It would be interesting to explore whether the models presented here appear differently when viewed through the $\epsilon$-machine framework~\cite{Shalizi2001}.

In computer science, it is sometimes possible to construct efficient deterministic algorithms from randomized ones by utilizing low entropy distributions.  One common technique is to replace the independent  binary random variables used in a~randomized algorithm with those satisfying only pairwise independence \cite{luby2006pairwise}.  In many cases, such a~randomized algorithm can be shown to succeed even if the original independent random bits are replaced by pairwise independent ones having significantly less entropy.  In particular, efficient derandomization can be accomplished in these instances by finding pairwise independent distributions with small sample spaces.  Several such designs are known and use tools from finite fields and linear codes \cite{joffe1974set, macwilliams1977error, luby86, alon1986fast, hedayat1999orthogonal}, combinatorial block designs \cite{hall1967combinatorial, karp1985fast}, Hadamard matrix theory \cite{lancaster1965pairwise, karloff1994construction}, and linear programming \cite{koller1993constructing}, among others \cite{Sylvester1867}.  Our construction here of two families of low entropy distributions fit to specified mean activities and pairwise statistics adds to this literature.

\section{Discussion}

Ideas and approaches from statistical mechanics are finding many applications in systems neuroscience~\cite{spikes,Madhu_Subhaneil_Surya_2013}.
In particular, maximum entropy models are powerful tools for understanding physical systems, and they are proving to be useful for describing biology as well, but a~deeper understanding of the full solution space is needed as we explore systems less amenable to arguments involving ergodicity or equally accessible states.
Here we have shown that second order statistics do not significantly constrain the range of allowed entropies, though other constraints, such as exchangeability, do guarantee extensive entropy (i.e., entropy proportional to system size $N$).

We have shown that in order for the constraints themselves to impose a~linear scaling on the entropy, the number of experimentally measured quantities that provide those constraints must scale exponentially with the size of the system.
In neuroscience, this is an unlikely scenario, suggesting that whatever means we use to infer probability distributions from the data (whether maximum entropy or otherwise) will most likely have to agree with other, more direct, estimates of the entropy~\cite{Panzeri2007, Rolls2011, Crumiller2011, Strong1998, Nemenman2004, Borst1999, QuianQuiroga2009}.
The~fact that maximum entropy models chosen to fit a~somewhat arbitrary selection of measured statistics are able to match the entropy of the system they model lends credence to the merits of this~approach.

Neural systems typically exhibit a~range of values for the correlations between pairs of neurons, with some firing coincidently more often than chance and others firing together less often than chance. Such systems can exhibit a~form of frustration, such that they apparently cannot be scaled up to arbitrarily large sizes in such a~way that the distribution of correlations and mean firing rates for the added neurons resembles that of the original system.  We have presented a~particularly simple example of a~small system with uniform pairwise correlations and mean firing rates that cannot be grown beyond a~specific finite size while maintaining these statistics throughout the network.

We have also indicated how, in some settings, minimum entropy models can provide a~floor on information transmission, complementary to channel capacity, which provides a~ceiling on system performance. Moreover, we show how highly entropic processes can be mimicked by low entropy processes with matching low-order statistics, which has applications in computer science.

%
%
%

%

\vspace{6pt}


\acknowledgments{The authors would like to thank Jonathan Landy, Tony Bell, Michael Berry, Bill Bialek, Amir~Khosrowshahi, Peter Latham, Lionel Levine, Fritz Sommer, and all members of the Redwood Center for many useful discussions.  M.R.D.\ is grateful for support from the Hellman Foundation, the McDonnell Foundation, the McKnight Foundation, the Mary Elizabeth Rennie Endowment for Epilepsy Research, and the National Science Foundation through Grant No. IIS-1219199.  C.H.\ was supported under an NSF All-Institutes Postdoctoral Fellowship administered by the Mathematical Sciences Research Institute through its core grant DMS-0441170.
This material is based upon work supported in part by the U.S. Army Research Laboratory and the U.S. Army Research Office under contract number W911NF-13-1-0390.}

\authorcontributions{Badr F. Albanna and Michael R. DeWeese contributed equally to this project and are resposible for the bulk of the results and writing of the paper. Jascha Sohl-Dickstein contributed substantially to clarifications of the main results and improvements to the text including the connection of these results to computer science.  Christopher Hillar also contributed substantially to clarifications of the main results, the~method used to identify an analytical lower bound on the minimum entropy solution, and improvements to the clarity of the text.  All authors approved of this manuscript before submission.}

\conflictsofinterest{The authors declare no conflict of interest.}

\appendixtitles{yes} 
\appendixsections{multiple} 
\appendix

\section{Allowed Range of \boldmath{$\nu$} Given \boldmath{$\mu$} Across All Distributions for Large \boldmath{$N$}}
\label{sec:range_of_mu_nu}

In this section we will only consider distributions satisfying uniform constraints
\begin{align}
    \mu_i &= \mu, \quad \forall i = 1,\ldots,N  \\
    \nu_{ij} &= \nu, \quad \forall i \neq j,
  \end{align}
and we will show that
\begin{equation}
\mu^2 \leq \nu \leq \mu
\end{equation}
in the large $N$ limit.
One could concievably extend the linear programming methods below to find bounds in the case of general non-uniform constraints, but as of this time we have not been able to do so without resorting to numerical algorithms on a~case-by-case basis.

We begin by determining the upper bound on $\nu$, the probability of any pair of neurons being simultaneously active, given $\mu$, the probability of any one neuron being active, in the large $N$ regime, where $N$ is the total number of neurons.
Time is discretized and we assume any neuron can spike no more than once in a~time bin.
We have $\nu \leq \mu$ because $\nu$ is the probability of a~pair of neurons firing together and thus each neuron in that pair must have at least a~firing probability of $\nu$.
Furthermore, it~is easy to see that the case $\mu$ = $\nu$ is feasible when there are only two states with non-zero probabilities: all neurons silent ($p_0$) or all neurons active ($p_1$).
In this case, $p_1 = \mu = \nu$. We use the term ``active'' to refer to neurons that are spiking, and thus equal to one, in a~given time bin, and we also refer to ``active'' states in a~distribution, which are those with non-zero probabilities.

We now proceed to show that the lower bound on $\nu$ in the large $N$ limit is $\mu^2$, the value of $\nu$ consistent with statistical independence among all $N$ neurons.
We can find the lower bound by viewing this as a~linear programming problem~\cite{Gale_1951,Boyd:2004:CO:993483}, where the goal is to maximize $-\nu$ given the normalization constraint and the constraints on $\mu$.

It will be useful to introduce the notion of an \textit{exchangeable distribution}~\cite{Diaconis_1977}, for which any permutation of the neurons in the binary words labeling the states leaves the probability of each state unaffected.
For example if $N=3$, an exchangeable solution satisfies
\begin{align}
p({100}) = p({010}) = p({001}), \nonumber \\
p({110}) = p({101}) = p({011}). \nonumber
\end{align}

In other words, the probability of any given word depends only on the number of ones it contains, not their particular locations, for an exchangeable distribution.

In order to find the allowed values of $\mu$ and $\nu$, we need only consider exchangeable distributions.  If there exists a~probability distribution that satisfies our constraints, we can always construct an~exchangeable one that also does given that the constraints themselves are symmetric (Equations~(\ref{eqn:min_summary}) and (\ref{eq:sym_mu})).
Let us do this explicitly:
Suppose we have a~probability distribution $p(\vec{s})$ over binary words $\vec{s} = (s_1,\ldots,s_N) \in \{0,1\}^N$ that satisfies our constraints but is not exchangeable.
We construct an~exchangeable distribution $p_e(w)$ with the same constraints as follows:
\begin{equation}
p_e(\vec{s}) \equiv \sum_{\sigma} \frac{p(\sigma(\vec{s}))}{N!},
\end{equation}
where $\sigma$ is an element of the permutation group $\mathcal{P}_N$ on $N$ elements.
This distribution is exchangeable by construction, and it is easy to verify that it satisfies the same uniform constraints as does the original distribution, $p(\vec{s})$.

Therefore, if we wish to find the maximum $-\nu$ for a~given value of $\mu$, it is sufficient to consider exchangeable distributions.
From now on in this section we will drop the $e$ subscript on our earlier notation, define $p$ to be exchangeable, and let $p(i)$ be the probability of a~state with $i$ spikes.

The normalization constraint is
\begin{equation}
1 = \sum_{i=0}^{N} \binom{N}{i} p(i).
\label{eqn:norm_constraint}
\end{equation}

Here the binomial coefficient $\binom{N}{i}$ counts the number of states with $i$ active neurons.

The firing rate constraint is similar, only now we must consider summing only those probabilities that have a~particular neuron active.
How many states are there with only a~pair of active neurons given that a~particular neuron must be active in all of the states?
We have the freedom to place the remaining active neuron in any of the $N-1$ remaining sites, which gives us $\binom{N-1}{1}$ states with probability $p(2)$.
In general if we consider states with $i$ active neurons, we will have the freedom to place $i-1$ of them in $N-1$ sites, yielding:
\begin{equation}
\mu = \sum_{i=1}^{N} \binom{N-1}{i-1} p(i) .
\label{eqn:mean_constraint}
\end{equation}

Finally, for the pairwise firing rate, we must add up states containing a~specific pair of active neurons, but the remaining $i-2$ active neurons can be anywhere else:
\begin{equation}
\nu = \sum_{i=2}^{N} \binom{N-2}{i-2} p(i) .
\label{eqn:pair_constraint}
\end{equation}

Now our task can be formalized as finding the maximum value of
\begin{equation}
-\nu = -\sum_{i=2}^{N} \binom{N-2}{i-2} p(i)
\end{equation}
subject to
\begin{align}
1		& = \sum_{i=0}^{N} \binom{N}{i} p(i), \\
\mu		& = \sum_{i=1}^{N} \binom{N-1}{i-1} p(i), \\
p(i)	& \geq 0, \quad \text{for all} \  i . \nonumber
\end{align}

This gives us the following dual problem:
Minimize
\begin{equation}
\label{eq:lambdas_and_mu}
\mathcal{E} \equiv \lambda_0 + \mu \lambda_1,
\end{equation}
given the following $N + 1$ constraints (each labeled by $i$)
\begin{equation}
\binom{N}{i} \lambda_0 + \binom{N-1}{i-1} \lambda_1  \geq - \binom{N-2}{i-2}, \quad N \geq i \geq 0,
\label{eqn:dual_constraints}
\end{equation}
where $\binom{a}{b}$ is taken to be zero for $b<0$.
The principle of strong duality~\cite{Boyd:2004:CO:993483} ensures that the value of the objective function at the solution is equal to the extremal value of the original objective function $-\nu$.

\begin{figure}[H]
\centering
\includegraphics[width=3.5in]{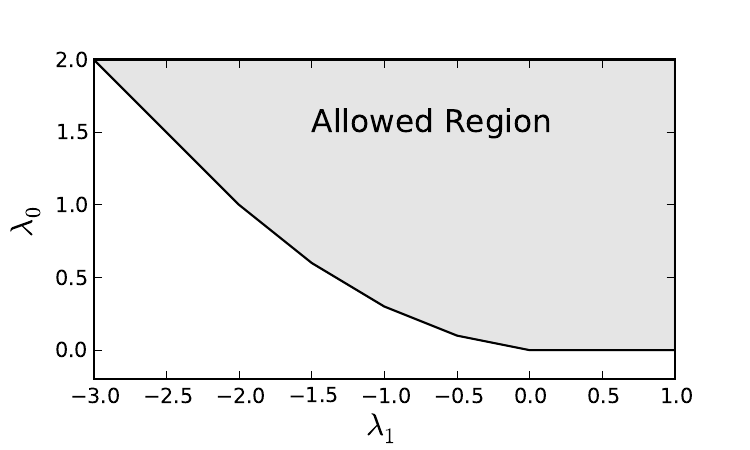}
\caption{An example of the allowed values of $\lambda_0$ and $\lambda_1$ for the dual problem ($N=5$).}
\label{fig:dual}
\end{figure}

The set of constraints defines a~convex region in the $\lambda_1$, $\lambda_0$ plane as seen in Figure \ref{fig:dual}.
The~minimum of our dual objective generically occurs at a~vertex of the boundary of the allowed region, or possibly degenerate minima can occur anywhere along an edge of the region.
From Figure~\ref{fig:dual} is is clear that this occurs where Equation~\eqref{eqn:dual_constraints} is an equality for two (or three in the degenerate case) consecutive values of $i$.
Calling the first of these two values $i_0$, we then have the following two equations that allow us to determine the optimal values of $\lambda_0$ and $\lambda_1$ ($\lambda^*_0$ and $\lambda^*_1$, respectively) as a~function of $i_0$
\begin{align}
 \binom{N}{i} \lambda_0^* + \binom{N-1}{i-1} \lambda_1^*  &= - \binom{N-2}{i_0-2} \\
 \binom{N}{i_0+1} \lambda_0^* + \binom{N-1}{i_0} \lambda_1^*  &= - \binom{N-2}{i_0-1}.
\end{align}

Solving for $\lambda^*_0$ and $\lambda^*_1$, we find
\begin{align}
  \lambda^*_0 &= \frac{i_0(i_0 +1)}{N(N-1)} \\
  \lambda^*_1 &= \frac{-2 i_0}{(N-1)}.
\end{align}

Plugging this into Equation \eqref{eq:lambdas_and_mu} we find the optimal value  $\mathcal{E}^*$ is
\begin{align}
  \mathcal{E}^* 	&= \lambda_0^* + \mu \lambda_1^* \nonumber \\
					&= \frac{i_0(i_0 +1)}{N(N-1)} - \mu  \frac{2 i_0}{(N-1)} \nonumber\\
					&= \frac{i_0 (i_0 + 1 - 2 \mu N)}{N(N-1)}.
  \label{eq:opt_dual}
\end{align}

Now all that is left is to express $i_0$ as a~function of $\mu$ and take the limit as $N$ becomes large.
This~expression can be found by noting from Equation~\eqref{eq:lambdas_and_mu} and Figure~\ref{fig:dual} that at the solution, $i_0$~satisfies

\begin{equation}
-m(i_0) \leq \mu \leq -m(i_0 + 1),
\label{eq:slope_mu}
\end{equation}
where $m(i)$ is the slope, $d\lambda_0/d\lambda_1$, of constraint $i$.
The expression for $m(i)$ is determined from Equation~\eqref{eqn:dual_constraints},
\begin{align}
m(i)	& = - \frac{\binom{N-1}{i-1}}{\binom{N}{i}} \nonumber \\
		& = - \frac{i}{N} .
		\label{eq:slope}
\end{align}

Substituting Equation \eqref{eq:slope} into Equation \eqref{eq:slope_mu}, we find
\begin{equation}
  \frac{i_0}{N} \leq \mu \leq  \frac{i_0+1}{N}.
\end{equation}

This allows us to write
\begin{equation}
  \mu = \frac{i_0 + b(N)}{N}
  \label{eq:mu_exp}.
\end{equation}
where $b(N)$ is between 0 and 1 for all $N$.
Solving this for $i_0$, we obtain
\begin{equation}
  i_0 = N \mu - b(N)
  \label{eq:i_0_exp}
\end{equation}

Substituting Equation~\eqref{eq:i_0_exp} into Equation~\eqref{eq:opt_dual}, we find
\begin{align}
  \mathcal{E}^* &= \frac{(N \mu - b(N))(N \mu - b(N) + 1 - 2 N\mu)}{N(N-1)}  \nonumber\\
				&= \frac{(N \mu - b(N))(- N \mu - b(N) + 1)}{N(N-1)} \nonumber \label{eq:mathcalE_finiteb}\\
				&= -\mu^2 + \mathcal{O}\left(\frac{1}{N}\right)
\end{align}

\begin{figure}[H]
\centering
\includegraphics[width=2in]{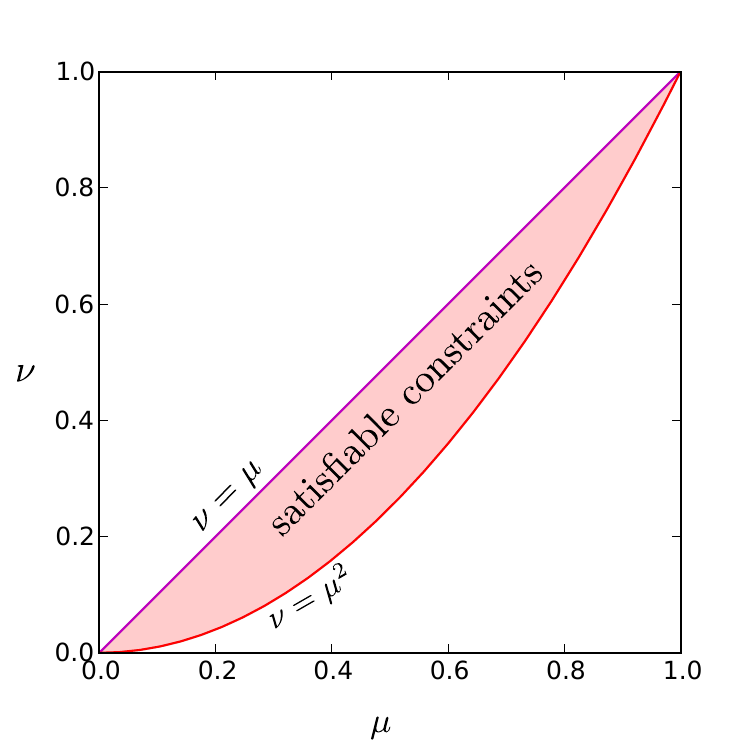}
\caption{The red shaded region is the set of values for $\mu$ and $\nu$ that can be satisfied for at least one probability distribution in the $N \rightarrow \infty$ limit. The purple line along the diagonal where $\nu = \mu$ is the distribution for which only the all active and all inactive states have non-zero probability. It represents the global entropy minimum for a~given value of $\mu$. The red parabola, $\nu = \mu^2$, at the bottom border of the allowed region corresponds to a~wide range of probability distributions, including the global maximum entropy solution for given $\mu$ in which each neuron fires independently. We find that low entropy solutions reside at this low $\nu$ boundary as well.}
\label{fig:allowed}
\end{figure}

Taking the large $N$ limit we find that $\mathcal{E}^* = -\mu^2$ and by the principle of strong duality ~\cite{Boyd:2004:CO:993483} the maximum value of $-\nu$ is $-\mu^2$.
Therefore we have shown that for large $N$, the region of satisfiable constraints is simply
\begin{equation}
\mu^2 \leq \nu \leq \mu,
\label{eqn:range_of_nu}
\end{equation}
as illustrated in Figure~\ref{fig:allowed}.

We can also compute upper and lower bounds on the minimum possible value for $\nu$ for finite $N$ by taking the derivative of $\mathcal{E}^*$ (Equation~(\ref{eq:mathcalE_finiteb})) with respect to $b(N)$ and setting that to zero, to obtain $b(N) = 0.5$. Recalling that $0 \le b(N) \le 1$, it is clear that the only candidates for extremizing $\mathcal{E}^*$ are $b(N) \in \{0, 0.5, 1\}$, and we have:
\begin{align}
\frac{N\mu(N\mu - 1)}{N(N-1)} \le \nu_{min}(N) \le \frac{(N\mu - 0.5)^2}{N(N-1)}.
\label{ineq:nu_min_bounds}
\end{align}

To obtain the exact value of the minimum of $\nu$ for finite $N$, we substitute the greatest integer less than or equal to $\mu N$ for $i_0$ in Equation~(\ref{eq:opt_dual}) to obtain
\begin{equation}
\nu_{min} = -\frac{\lfloor \mu N\rfloor\left( \lfloor\mu N\rfloor +  1 - 2 \mu N\right)}{N (N-1)},
\end{equation}
where $\lfloor . \rfloor$ is the floor function.
Both of the bounds in (\ref{ineq:nu_min_bounds})
 and the true $\nu_{min}$ are plotted as functions of $N$ in Figure~\ref{fig:crash} of the main text for $\mu = 0.1$.

\section{Minimum Entropy Occurs at Small Support}
\label{sec:support}

Our goal is to minimize the entropy function
\begin{equation}
S(\mathbf{p}) = \sum_{i=0}^{n_s} - p_i \log_2 p_i,
\end{equation}
where $n_s$ is the number of states, the $p_i$ satisfy a~set of $n_c$ independent linear constraints, and $p_i \geq 0$ for all $i$.  For the main problem we consider, $n_s = 2^N$.  The constraints for normalization, mean firing rates, and pairwise firing rates give
\begin{align}
n_c & = 1 + N + \frac{N(N-1)}{2}  \nonumber\\
    & = 1 + \frac{N(N+1)}{2}.
\end{align}

In this section we will show that the minimum occurs at the vertices of the space of allowed probabilities.
Moreover, these vertices correspond to probabilities of small support---specifically a~support size equal to $n_c$ in most cases.
These two facts allow us to put an upper bound on the minimum entropy of
\begin{equation}
\tilde{S}_2 \leq \tilde{S}_2^{hi} \approx 2 \log_2 (N),
\end{equation}
for large $N$.

We begin by noting that the space of normalized probability distributions \mbox{$\mathcal{P}=\{p:\sum_{i \in 1}^{n_s} {p_i}~=~1, \ p_i \geq 0\}$} is
the standard simplex in $n_s - 1$ dimensions.
Each linear constraint on the probabilities introduces a~hyperplane in this space.  If the constraints are consistent and independent, the intersection of these hyperplanes defines a~$d = n_s - n_c$ affine space, which we call $\mathcal{C}$.
All solutions are constrained to the intersection between $\mathcal{P}$ and $\mathcal{C}$ and this solution space is a~convex polytope of dimension
$\le$~$d$, which we refer to as $\mathcal{R}$.
A point within a~convex polytope can always be expressed as a~linear combination of its vertices, therefore if $\{\mathbf{v}_i\}$ are the vertices of $\mathcal{R}$
we may write
\begin{equation}
\mathbf{p} = \sum_i^{n_v} a_i \mathbf{v}_i,
\end{equation}
where $n_v$ is the total number of vertices and $\sum_i^{n_v} a_i = 1$.

Using the concavity of the entropy function, we will now show that the minimum entropy for a~space of probabilities $S$ is attained on the vertices of that space.  Of course, this means that the global minimum will occur at the vertex that has the lowest entropy, $\mathbf{v}_*$
\begin{align}
S(\mathbf{p}) & = S\left(\sum_i^{n_v} a_i \mathbf{v}_i\right)  \nonumber\\
              & \geq \sum_i^{v_s} a_i S(\mathbf{v}_i)  \nonumber\\
              & \geq \left( \sum_i^{v_s} a_i \right) S(\mathbf{v}_*)  \nonumber\\
              & = S(\mathbf{v}_*).
\end{align}

Therefore,
\begin{equation}
\tilde{S}_2 = S(\mathbf{v}_*).
\end{equation}

Moreover, if a~distribution satisfying the constraints exists, then there is one with at most $n_c$ non-zero $p_i$ ({e.g.}, from arguments as in \cite{koller1993constructing}).  Together, these two facts imply that there are minimum entropy distributions with a~maximum of $n_c$ non-zero $p_i$.  This means that even though the state space may grow exponentially with $N$, the support of the minimum entropy solution for fixed means and pairwise correlations will only scale quadratically with $N$.

This allows us to give an upper bound on the minimum entropy as,
\begin{align}
\tilde{S}_2 \leq \tilde{S}_2^{hi}   & = \log_2(n_c) \nonumber \\
                                    & = \log_2\left(1 + \frac{N(N+1)}{2}\right)  \nonumber\\
                                    & \approx 2 \log_2 (N),
\end{align}
for large $N$.
It is important to note how general this bound is: as long as the constraints are independant and consistent this result holds \emph{regardless} of the specific values of the $\{\mu_i\}$ and $\{\nu_{ij}\}$.

\section{The Maximum Entropy Solution}
\label{sec:max_entropy}

In the previous Appendix, we derived a~useful upper bound on the minimum entropy solution valid for any values of $\{\mu_i\}$ and $\{\nu_{ij}\}$ that can be achieved by at least one probability distribution.
In Appendix~\ref{sec:lower_bound} below, we obtain a~useful lower bound on the minimum entropy solution.
It is straightforward to obtain an upper bound on the maximum entropy distribution valid for arbitrary achievable $\{\mu_i\}$ and $\{\nu_{ij}\}$:
the greatest possible entropy for $N$ neurons is achieved if they all fire independently with probability $1/2$, resulting in the bound $S \leq N$.

Deriving a~useful lower bound for the maximum entropy for arbitrary allowed constraints $\{\mu_i\}$ and $\{\nu_{ij}\}$ is a~subtle problem.
In fact, merely specifying how an ensemble of binary units should be grown from some finite initial size to arbitrarily large $N$ in such a~way as to ``maintain'' the low-level statistics of the original system raises many questions.

For example, typical neural populations consist of units with varying mean activities, so how should the mean activities of newly added neurons be chosen?  For that matter, what correlational structure should be imposed among these added neurons and between each new neuron and the existing units?  For each added neuron, any choice will inevitably change the histograms of mean activities and pairwise correlations for any initial ensemble consisting of more than one neuron, except in the special case of uniform constraints.  Even for the relatively simple case of uniform constraints, we have seen that there are small systems that cannot be grown beyond some critical size while maintaining uniform values for $\{\mu\}$ and $\{\nu\}$ (see Figure~2 in the main text). The problem of determining whether it is even mathematically possible to add a~neuron with some predetermined set of pairwise correlations with existing neurons can be much more challenging for the general case of nonuniform constraints.

For these reasons, we will leave the general problem of finding a~lower bound on the maximum entropy for future work and focus here on the special case of uniform constraints:
\begin{align}
  \label{eqn:sym_mu}
    \mu_i &= \mu, \quad \forall i = 1,\ldots,N\\
  \label{eqn:sym_nu}
    \nu_{ij} &= \nu, \quad \forall i \neq j.
\end{align}

We will obtain a~lower bound on the maximum entropy of the system with the use of an explicit construction, which will necessarily have an entropy, $S^{con}_2$, that does not exceed that of the maximum entropy solution.  We construct our model system as follows: with probability $\beta$, all $N$ neurons are active and set to 1, otherwise each neuron is independently set to 1 with probability $\eta$.  The conditions required for this model to match the desired mean activities and pairwise correlations across the population are given by
\begin{align}
  \label{eqn:con_3_mu}
    \mu &= \beta + (1 - \beta)\eta,\\
  \label{eqn:con_3_nu}
    \nu &= \beta + (1 - \beta)\eta^2.
\end{align}

Inverting these equations to isolate $\beta$ and $\eta$ yields
\begin{align}
  \label{eqn:con_3_beta}
    \beta &= \frac{\nu - \mu^2}{1 + \nu - 2\mu},\\
  \label{eqn:con_3_eta}
    \eta &= \frac{\mu - \nu}{1 - \mu}.
\end{align}

The entropy of the system is then
\begin{align}
  \label{eqn:S_con_3}
    S^{con}_2 = & -\beta \log_2\beta - (1 - \beta)\log_2 (1 - \beta)  \cr
                         & + (1-\beta) N \left[-\eta\log_2\eta - (1 - \eta)\log_2(1 - \eta)\right] \cr
                         = & \; w + xN,
\end{align}
where $w$ and $x$ are nonnegative constants for all allowed values of $\mu$ and $\nu$ that can be achieved for arbitrarily large systems. $x$ is nonzero provided $\nu < \mu$ ({ i.e.}, $\beta \neq 1 \cap \eta \neq \{0,1\}$), so the entropy of the system will grow linearly with $N$ for any uniform constraints achievable for arbitrarily large systems except for the case in which all neurons are perfectly correlated.

Using numerical methods, we can empirically confirm the linear scaling of the entropy of the true maximum entropy solution for the case of uniform constraints that can be achieved by arbitrarily large systems.
In general, the constraints can be written

\begin{align}
\mu_i	& = \avg{s_i} = \sum_{\vec{s}} p(\vec{s}) s_i, \quad i = 1,\ldots,N,  \label{eqn:const_1} \\
\nu_{ij}		& = \avg{s_i s_j} = \sum_{\vec{s}} p(\vec{s}) s_i s_j, \quad i \neq j, \label{eqn:const_2}
\end{align}
where the sums run over all $2^N$ states of the system.
In order to enforce the constraints, we can add terms involving Lagrange multipliers $\lambda_i$ and $\gamma_{ij}$ to the entropy in the usual fashion to arrive at a~function to be maximized
\begin{equation}
\begin{split}
\mathcal{S}(p(\vec{s})) = & -\sum_{\vec{s}} p(\vec{s}) \log_2 p(\vec{s}) \\
		& {} - \sum_i \lambda_i \left( \sum_{\vec{s}} p(\vec{s}) s_i - \mu_i \right) \\
		& {} - \sum_{i<j} \gamma_{ij} \left( \sum_{\vec{s}} p(\vec{s}) s_i s_j - \nu_{ij} \right).
\end{split}
\label{eqn:entropy_objective}
\end{equation}

Maximizing this function gives us the Boltzmann distribution for an Ising spin glass
\begin{equation}
p(\vec{s}) = \frac{1}{\mathcal{Z}} \exp\left(-\sum_i \lambda_i s_i - \sum_{i<j} \gamma_{ij} s_i s_j \right),
\label{eq:max_ent_prob}
\end{equation}
where $\mathcal{Z}$ is the normalization factor or partition function.
The values of $\lambda_i$ and $\gamma_{ij}$ are left to be determined by ensuring this distribution is consistent with our constraints $\{\mu_i\}$ and $\{\nu_{ij}\}$.

For the case of uniform constraints, the Lagrange multipliers are themselves uniform:
\begin{align}
\lambda_i & = \lambda, \quad \forall i ,\\
\gamma_{ij} & = \gamma, \quad \forall i < j.
\end{align}
This allows us to write the following expression for the maximum entropy distribution:
\begin{equation}
p(\vec{s}) = \frac{1}{\mathcal{Z}} \exp\left(-\lambda \sum_i s_i - \gamma \sum_{i<j} s_i s_j \right).
\end{equation}
If there are $k$ neurons active, this becomes
\begin{equation}
p(k) = \frac{1}{\mathcal{Z}} \exp\left(-\lambda k - \gamma \frac{k(k-1)}{2} \right) \label{eqn:form_of_p}.
\end{equation}
Note that there are $\binom{N}{k}$ states with probability $p(k)$.
Using expression \eqref{eqn:form_of_p}, we find the maximum entropy by using the \texttt{fsolve} function from the \texttt{SciPy} package of Python subject to constraints \eqref{eqn:sym_mu} and \eqref{eqn:sym_nu}.

\begin{figure}[H]
\centering
\includegraphics[width=2.3in]{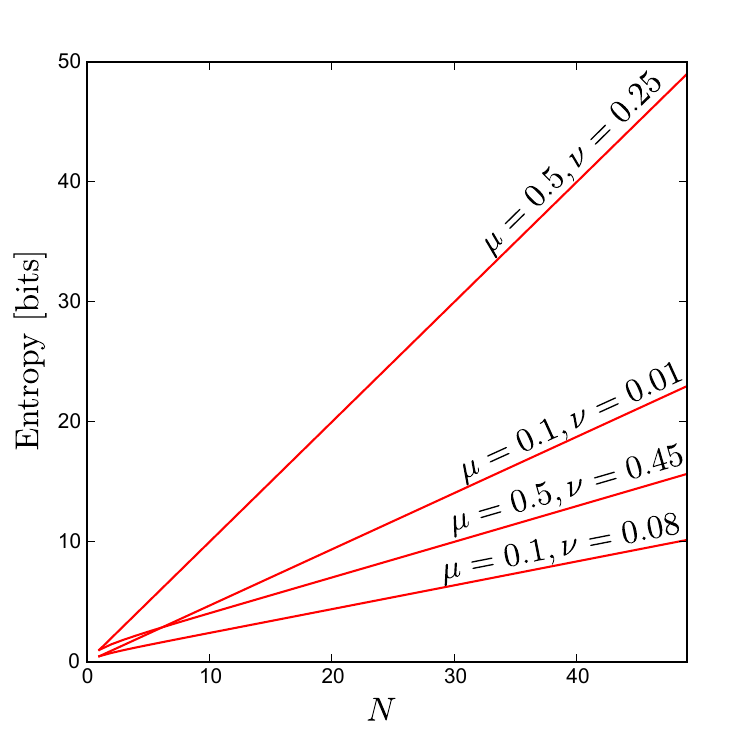}
\caption{For the case of uniform constraints
achievable by arbitrarily large systems,
the maximum possible entropy scales linearly with system size, $N$, as shown here for various values of $\mu$ and $\nu$. Note~that this linear scaling holds even for large correlations,
provided that $\nu < \mu$. For the extreme case $\nu = \mu$, all the neurons are perfectly correlated so the entropy of the ensemble does not change with increasing $N$.
}
\label{fig:max}
\end{figure}

As Figure~\ref{fig:max} shows, for fixed uniform constraints the entropy scales linearly as a~function of $N$, even in cases where the correlations between all pairs of neurons ($\nu$) are quite large, provided that $\mu^2 \leq \nu < \mu$.
Moreover, for uniform constraints with anticorrelated units ({ i.e.}, pairwise correlations {\em below} the level one would observe for independent units), we find empirically that the maximum entropy still scales approximately linearly up to the maximum possible system size consistent with these constraints, as illustrated by Figure~\ref{fig:crash} in the main text.

\section{Minimum Entropy for Exchangeable Probability Distributions}
\label{sec:exch_entropy}

Although the values of the firing rate ($\mu$) and pairwise correlations ($\nu$) may be identical for each neuron and pair of neurons, the probability distribution that gives rise to these statistics need not be exchangeable as we have already shown.
Indeed, as we explain below, it is possible to construct non-exchangeable probability distributions that have dramatically lower entropy then both the maximum and the minimum entropy for exchangeable distributions.
That said, exchangeable solutions are interesting in their own right because they have large $N$ scaling behavior that is distinct from the global entropy minimum, and they provide a~symmetry that can be used to lower bound the information transmission rate close to the maximum possible across all distributions.

Restricting ourselves to exchangeable solutions represents a~significant simplification.
In the general case, there are $2^N$ probabilities to consider for a~system of $N$ neurons.
There are $N$ constraints on the firing rates (one for each neuron) and $\binom{N}{2}$ pairwise constraints (one for each pair of neurons).
This gives us a~total number of constrains ($n_c$) that grows quadratically with $N$:
\begin{equation}
n_c = 1 + \frac{N(N+1)}{2}.
\end{equation}

However in the exchangeable case, all states with the same number of spikes have the same probability so there are only $N+1$ free parameters.
Moreover, the number of constraints becomes 3 as there is only one constraint each for normalization, firing rate, and pairwise firing rate (as expressed in Equations~\eqref{eqn:norm_constraint}--\eqref{eqn:pair_constraint}, respectively).
In general, the minimum entropy solution for exchangeable distributions should have the minimum support consistent with these three constraints.
Therefore, the minimum entropy solution should have at most three non-zero probabilities (see Appendix~\ref{sec:support}).

For the highly symmetrical case with $\mu = \nicefrac{1}{2}$ and $\nu = \nicefrac{1}{4}$, we can construct the exchangeable distribution with minimum entropy for all even $N$.
This distribution consists of the all ones state, the~all zeros state, and all states with $N/2$ ones. The constraint $\mu = \nicefrac{1}{2}$ implies that $p(0) = p(N)$, and~the condition $\nu = \nicefrac{1}{4}$ implies
\begin{equation}
p(N/2) = \frac{N - 1}{N}\frac{(N/2)!^2}{N!}, \quad N \textrm{ even},
\end{equation}
which corresponds to an entropy of
\begin{equation}
\tilde{S}_2^{exch} = \frac{\log_2(2N)}{N} + \frac{N-1}{N} \log_2\left(\frac{NN!}{(N/2)!^2(N-1)}\right)  \label{eq:exch__sym_min}.
\end{equation}

{Using Sterling's approximation and taking the large $N$ limit, this simplifies to}
\begin{equation}
\tilde{S}_2^{exch} \approx N - \nicefrac{1}{2} \log_2(N) - \nicefrac{1}{2} \log_2(2 \pi) + O\left[\frac{\log_2(N)}{N}\right]. \label{eq:exch__sym_min_approx}
\end{equation}

For arbitrary values of $\mu$, $\nu$ and $N$, it is difficult to determine from first principles which three probabilities are non-zero for the minimum entropy solution, but fortunately the number of possibilities $\binom{N+1}{3}$ is now small enough that we can exhaustively search by computer to find the set of non-zero probabilities corresponding to the lowest entropy.

\begin{figure}[H]
\centering
\includegraphics[width=2.3in]{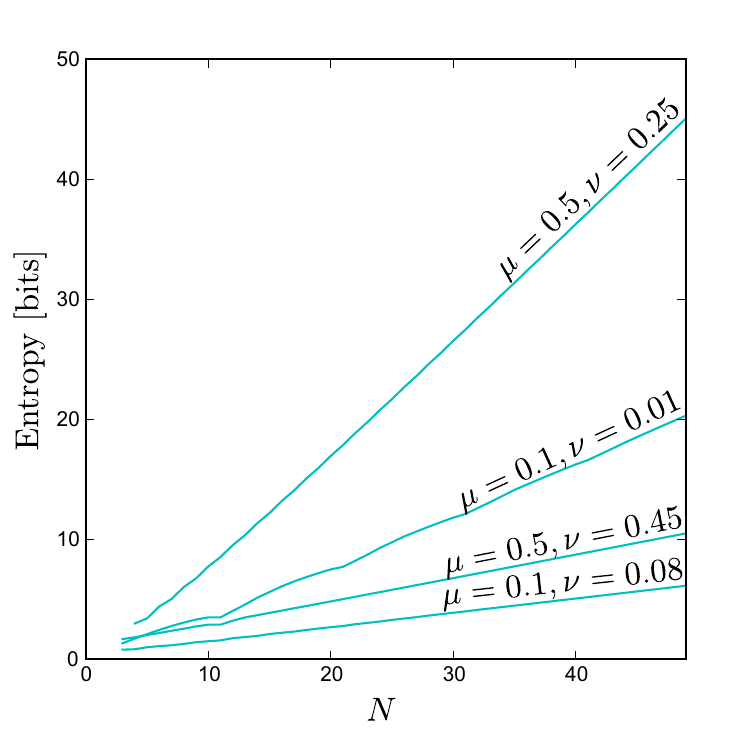}
\caption{The minimum entropy for exchangeable distributions versus $N$ for various values of $\mu$ and $\nu$. Note that, like the maximum entropy, the exchangeable minimum entropy scales linearly with $N$ as $N \rightarrow \infty$, albeit with a~smaller slope for $\nu \neq \mu^2$. We can calculate the entropy exactly for $\mu$ = 0.5 and $\nu$ = 0.25 as $N \rightarrow \infty$, and we find that the leading term is indeed linear:
\mbox{$\tilde{S}_2^{exch} \approx N - 1/2 \log_2(N) - 1/2 \log_2(2 \pi) + O[\log_2(N)/N]$}.}
\label{fig:exch_min}
\end{figure}

Using this technique, we find that the scaling behavior of the exchangeable minimum entropy is linear with $N$ as shown in Figure~\ref{fig:exch_min}.
We find that the asymptotic slope is positive, but less than that of the maximum entropy curve, for all $\nu \neq \mu^2$.  For the symmetrical case, $\nu = \mu^2$, our exact expression Equation~\eqref{eq:exch__sym_min} for the exchangeable distribution consisting of the all ones state, the all zeros state, and~all states with $N/2$ ones agrees with the minimum entropy exchangeable solution found by exhaustive search, and in this special case the asymptotic slope is identical to that of the maximum entropy curve.

\section{Construction of a~Low Entropy Distribution for All Values of \boldmath{$\mu$} and \boldmath{$\nu$}}
\label{sec:prime_construction}

We can construct a~probability distribution with roughly $N^2$ states with nonzero probability out of the full $2^N$ possible states of the system such that
\begin{equation}
\mu = \frac{n}{N}, \quad \nu = \frac{n(n-1)}{N(N-1)},
\label{Const:values}
\end{equation}
where $N$ is the number of neurons in the network and $n$ is the number of neurons that are active in every state.
Using this solution as a~basis, we can include the states with all neurons active and all neurons inactive to create a~low entropy solution for all allowed values for $\mu$ and $\nu$ (See Appendix~\ref{sec:validity}).  We refer to the entropy of this low entropy construction $\tilde{S}^{con2}_2$ to distinguish it from the entropy ($\tilde{S}^{con}_2$) of another low entropy solution described in the next section.  Our construction essentially goes back to Joffe~\cite{joffe1974set} as explained by Luby in \cite{luby86}.

We derive our construction by first assuming that $N$ is a~prime number,
but this is not actually a~limitation as we will be able to extend the result to all values of $N$.
Specifically, non-prime system sizes are handled by taking a~solution for a~larger prime number and removing the appropriate number of neurons.
It should be noted that occasionally the solution derived using the next largest prime number does not necessarily have the lowest entropy and occasionally we must use even larger primes to find the minimum entropy possible using this technique; all plots in the main text were obtained by searching for the lowest entropy solution using the 10 smallest primes that are each at least as great as the system size $N$.

We begin by illustrating our algorithm with a~concrete example; following this illustrative case we will prove that each step does what we expect in general.
Consider $N=5$, and $n=3$.
The algorithm is as follows:
\begin{enumerate}
\item Begin with the state with $n=3$ active neurons in a~row:

\begin{tabular}{c}
11100
\end{tabular}

\item Generate new states by inserting progressively larger gaps of 0s before each 1 and wrapping active states that go beyond the last neuron back to the beginning.  This yields $N-1=4$ unique states including the original state:

\begin{tabular}{c}
11100 \\
10101 \\
11010 \\
10011 \\
\end{tabular}

\item Finally, ``rotate'' each state by shifting each pattern of ones and zeros to the right (again wrapping states that go beyond the last neuron). This yields a~total of $N(N-1)$ states:

\begin{tabular}{c c c c c}
11100 & 01110 & 00111 & 10011 & 11001 \\
10101 & 11010 & 01101 & 10110 & 01011 \\
11010 & 01101 & 10110 & 01011 & 10101 \\
10011 & 11001 & 11100 & 01110 & 00111 \\
\end{tabular}

\item Note that each state is represented twice in this collection, removing duplicates we are left with $N(N-1)/2$ total states.  By inspection we can verify that each neuron is active in $n(N-1)/2$ states and each pair of neurons is represented in $n(n-1)/2$ states.  {Weighting} each state with equal probability gives us the values for $\mu$ and $\nu$ stated in Equation~\eqref{Const:values}.
\end{enumerate}

Now we will prove that this construction works in general for $N$ prime and any value of $n$ by establishing (1) that step 2 of the above algorithm produces a~set of states with $n$ spikes;
(2) that this method produces a~set of states that when weighted with equal probability yield neurons that all have the same firing rates and pairwise statistics; and (3) that this method produces at least double redundancy in the states generated as stated in step 4 (although in general there may be a~greater redundancy).
In discussing (1) and (2) we will neglect the issue of redundancy and consider the states produced through step 3 as distinct.

First we prove that step 2 always produces states with $n$ neurons, which is to say that no two spikes are mapped to the same location as we shift them around.
We will refer to the identity of the spikes by their location in the original starting state; this is important as the operations in step 2 and 3 will change the relative ordering of the original spikes in their new states.
With this in mind, the~location of the $i$th spike with a~spacing of $s$ between them will result in the new location $l$ (here the original state with all spikes in a~row is $s=1$):
\begin{equation}
l = (s \cdot i) \bmod N  ,
\end{equation}
where $i \in \{0,1,2,...,n-1\}$.
In this form, our statement of the problem reduces to demonstrating that for given values of $s$ and $N$, no two values of $i$ will result in the same $l$.
This is easy to show by contradiction.
If this were the case,
\begin{eqnarray}
(s \cdot i_1) \bmod N =  (s \cdot i_2) \bmod N \nonumber \\
\Rightarrow (s \cdot (i_1-i_2)) \bmod N = 0 .
\end{eqnarray}

For this to be true, either $s$ or $(i_1-i_2)$ must contain a~factor of $N$, but each are smaller than $N$ so we have a~contradiction.
This also demonstrates why $N$ must be prime---if it were not, it would be possible to satisfy this equation in cases where $s$ and $(i_1-i_2)$ contain between them all the factors of~$N$.

It is worth noting that this also shows that there is a~one-to-one mapping between $s$ and $l$ given $i$.
In other words, each spike is taken to every possible neuron in step 2.
For example, if $N=5$, and we fix $i=2$:
\begin{eqnarray*}
0 \cdot 2 \bmod 5 & = & 0 \\
1 \cdot 2 \bmod 5 & = & 2 \\
2 \cdot 2 \bmod 5 & = & 4 \\
3 \cdot 2 \bmod 5 & = & 1 \\
4 \cdot 2 \bmod 5 & = & 3
\end{eqnarray*}

If we now perform the operation in step 3, then the location $l$ of spike $i$ becomes
\begin{equation}
l = (s \cdot i + d ) \bmod N \label{Const:mathrep},
\end{equation}
where $d$ is the amount by which the state has been rotated (the first column in step 3 is $d=0$, the~second is $d=1$, etc.).
It should be noted that step 3 trivially preserves the number of spikes in our states so we have established that steps 2 and 3 produce only states with $n$ spikes.

We now show that each neuron is active, and each pair of neurons is simultaneously active, in the same number of states.
This way when each of these states is weighted with equal probability, we find symmetric statistics for these two quantities.

Beginning with the firing rate, we ask how many states contain a~spike at location $l$.
In other words, how many combinations of $s$, $i$, and $d$ can we take such that Equation~\eqref{Const:mathrep} is satisfied for a~given $l$.
For each choice of $s$ and $i$ there is a~unique value of $d$ that satisfies the equation.
$s$ can take values between $1$ and $N-1$, and $i$ takes values from $0$ to $n-1$, which gives us
$n(N-1)$ states that include a~spike at location $l$.
Dividing by the total number of states $N(N-1)$ we obtain an average firing rate of
\begin{equation}
\mu = \frac{n}{N}.
\end{equation}

Consider neurons at $l_1$ and $l_2$; we wish to know how many values of $s$, $d$, $i_1$ and $i_2$ we can pick so~that
\begin{eqnarray}
l_1 & = & (s \cdot i_1 + d ) \bmod N \label{Const:pairloc1}, \\
l_2 & = & (s \cdot i_2 + d ) \bmod N \label{Const:pairloc2}.
\end{eqnarray}

Taking the difference between these two equations, we find
\begin{equation}
\Delta l = (s \cdot (i_2 - i_1) ) \bmod N.
\end{equation}

From our discussion above, we know that this equation uniquely specifies $s$ for any choice of $i_1$ and $i_2$.
Furthermore, we must pick $d$ such that Equations~\eqref{Const:pairloc1} and  \eqref{Const:pairloc2} are satisfied.
This means that for each choice of $i_1$ and $i_2$ there is a~unique choice of $s$ and $d$, which results in a~state that includes active neurons at locations $l_1$ and $l_2$.
Swapping $i_1$ and $i_2$ will result in a~different $s$ and $d$.
Therefore, we have $n(n-1)$ states that include any given pair---one for each choice of $i_1$ and $i_2$.
Dividing this number by the total number of states, we find a~correlation $\nu$ equal to
\begin{equation}
\nu = \frac{n(n-1)}{N(N-1)},
\end{equation}
where $N$ is prime.

Finally we return to the question of redundancy among states generated by steps 1 through 3 of the algorithm.
Although in general there may be a~high level of redundancy for choices of $n$ that are small or close to $N$, we can show that in general there is at least a~twofold degeneracy.
Although this does not impact our calculation of $\mu$ and $\nu$ above, it does alter the number of states, which will affect the entropy of system.

The source of the twofold symmetry can be seen immediately by noting that the third and fourth rows of our example contain the same set of states as the second and first respectively.
The reason for this is that each state in the $s=4$ case involves spikes that are one leftward step away from each other just as $s=1$ involves spikes that are one rightward shift away from each other.
The labels we have been using to refer to the spikes have reversed order but the set of states are identical.
Similarly the $s=3$ case contains all states with spikes separated by two leftward shifts just as the $s=2$ case.
Therefore, the set of states with $s=a$ is equivalent to the set of states with $s = N-a$.
Taking this degeneracy into account, there are at most $N(N-1)/2$ unique states; each neuron spikes in $n(N-1)/2$ of these states and any given pair spikes together in $n(n-1)/2$ states.

Because these states each have equal probability the entropy of this system is bounded from above~by
\begin{equation}
\tilde{S}^{con2}_2 \leq \log_2 \left(\frac{N(N-1)}{2}\right),  
\end{equation}
where $N$ is prime.
As mentioned above, we write this as an inequality because further degeneracies among states beyond the factor of two that always occurs are possible for some prime numbers.  In~fact, in order to avoid non-monotonic behavior, the curves for $S^{con2}_2$
shown in Figures \ref{fig:SvN} and \ref{fig:crash} of the main text were generated using the lowest entropy found for the 10 smallest primes greater than $N$ for each value of $N$.

We can extend this result to arbitrary values for $N$ including non-primes by invoking the Bertrand-Chebyshev theorem, which states that there always exists at least one prime number $p$ with $n < p < 2n - 2$ for any integer $n > 1$:
\begin{equation}
\tilde{S}^{con2}_2 \leq \log_2 \left(N(2N-1)\right),
\end{equation}
where $N$ is any integer.
Unlike the maximum entropy and the entropy of the exchangeable solution, which we have shown to both be extensive quantities, this scales only logarithmically with the system size $N$.

\section{Another Low Entropy Construction for the Communications Regime, \boldmath{$\mu = \nicefrac{1}{2}$} \& \boldmath{$\nu = \nicefrac{1}{4}$}}
\label{sec:comm_construction}

In addition to the probability distribution described in the previous section, we also rediscovered another low entropy construction in the regime most relevant for engineered communications systems ($\mu = \nicefrac{1}{2}$, $\nu = \nicefrac{1}{4}$) that allows us to satisfy our constraints for a~system of $N$ neurons with only $2N$ active states.
Below we describe a~recursive algorithm for determining the states for arbitrarily large systems---states needed for $N=2^q$ are built from the states needed for $N = 2^{q-1}$, where $q$ is any integer greater than 2.
This is sometimes referred to as a~Hadamard matrix. Interestingly, this specific example goes back to Sylvester in 1867~\cite{Sylvester1867}, and it was recently discussed in the context of neural modeling by Macke and colleagues~\cite{Macke2011}.

We begin with $N=2^1=2$.
Here we can easily write down a~set of states that when weighted equally lead to the desired statistics.  Listing these states as rows of zeros  and ones, we see that they include all possible two-neuron states:
\begin{equation}
\begin{array}{cc}
1 & 1 \\
0 & 1 \\
0 & 0 \\
1 & 0
\label{eqn:2_neuron_mike}
\end{array}
\end{equation}

In order to find the states needed for $N=2^2=4$ we replace each $1$ in the above by
\begin{equation}
\begin{array}{cc}
1 & 1 \\
0 & 1
\end{array}
\label{eqn:one_sub}
\end{equation}
and each $0$ by
\begin{equation}
\begin{array}{cc}
0 & 0 \\
1 & 0
\end{array}
\label{eqn:zero_sub}
\end{equation}
to arrive at a~new array for twice as many neurons and twice as many states with nonzero probability:
\begin{equation}
\begin{array}{cccc}
1 & 1 & 1 & 1 \\
0 & 1 & 0 & 1 \\
0 & 0 & 1 & 1 \\
1 & 0 & 0 & 1 \\
0 & 0 & 0 & 0 \\
1 & 0 & 1 & 0 \\
1 & 1 & 0 & 0 \\
0 & 1 & 1 & 0
\end{array}
\end{equation}

By inspection, we can verify that each new neuron is spiking in half of the above states and each pair is spiking in a~quarter of the above states.
This procedure preserves $\mu=\nicefrac{1}{2}$, $\nu=\nicefrac{1}{4}$, and~\mbox{$\avg{s_i s_j s_k} = \nicefrac{1}{8}$} for all neurons; thus providing a~distribution that mimics the statistics of independent binary variables up to third order (although it does not for higher orders).
Let us consider the the proof that $\mu = \nicefrac{1}{2}$ is preserved by this transformation.
In the process of doubling the number of states from $N^q$ to $N^{q+1}$, each neuron with firing rate $\mu^{(q)}$ ``produces'' two new neurons with firing rates $\mu_1^{(q+1)}$ and $\mu_2^{(q+1)}$.
It is clear from Equations~\eqref{eqn:one_sub} and \eqref{eqn:zero_sub} that we obtain the following two~relations,
\begin{align}
	\mu_1^{(q+1)} &= \mu^{(q)}, \\
	\mu_2^{(q+1)} &= \nicefrac{1}{2}.
\end{align}

It is clear from these equations that if we begin with $\mu^{(1)} = \nicefrac{1}{2}$ that this will be preserved by this transformation.
By similar, but more tedious, methods one can show that $\nu=\nicefrac{1}{4}$, and $\avg{s_i s_j s_k} = \nicefrac{1}{8}$.

Therefore, we are able to build up arbitrarily large groups of neurons that satisfy our statistics using only $2N$ states by repeating the procedure that took us from $N=2$ to $N=4$.
Since these states are weighted with equal probability we have an entropy that grows only logarithmically with $N$
\begin{equation}
\label{eq:con_for_power_two}
  \tilde{S}^{con}_2 = \log_2 (2N), \quad N = 2^q, \  q = 2,3,4,\ldots.
\end{equation}

We mention briefly a~geometrical interpretation of this probability distribution.  The active states in this distribution can be thought of as a~subset of $2N$ corners on an $N$ dimensional hypercube with the property that the separation of almost every pair is the same. Specifically, for each active state, all~but one of the other active states has a~Hamming distance of exactly $N/2$ from the original state; the~remaining state is on the opposite side of the cube, and thus has a~Hamming distance of $N$.  In other words, for any pair of polar opposite active states, there are $2N-2$ active states around the ``equator''.

We can extend Equation~(\ref{eq:con_for_power_two}) to arbitrary numbers of neurons that are not multiples of 2 by taking the least multiple of 2 at least as great as $N$, so that in general:
\begin{equation}
\tilde{S}^{con}_2 = \lceil \log_2(2N)\rceil \leq \log_2 (N) + 2, \quad N \geq 2.
\label{eq:S_con}
\end{equation}

By adding two other states we can extend this probability distribution so that it covers most of the allowed region for $\mu$ and $\nu$ while remaining a~low entropy solution, as we now describe.

We remark that the authors of~\cite{chor1985bit, alon1986fast} provide a~lower bound of {$\Omega_N$} for the sample size possible for a~pairwise independent binary distribution, making the sample size of our novel construction essentially optimal.

\section{Extending the Range of Validity for the Constructions}
\label{sec:validity}

We now show that each of these low entropy probability distributions can be generalized to cover much of the allowed region depicted in Figure~\ref{fig:allowed}; in fact, the distribution derived in Appendix~\ref{sec:prime_construction} can be extended to include all possible combinations of the constraints $\mu$ and $\nu$.
This can be accomplished by including two additional states: the state where all neurons are silent and the state where all neurons are active.
If we weight these states by probabilities $p_0$ and $p_1$ respectively and allow the $N(N-1)/2$ original states to carry probability $p_n$ in total, normalization requires
\begin{equation}
p_0 + p_n + p_1 = 1.
\end{equation}

We can express the value of the new constraints  ($\mu'$ and $\nu'$) in terms of the original constraint values ($\mu$ and $\nu$) as follows:
\begin{align}
\mu'	& = (1 - p_0 - p_1) \mu + p_1 \label{eqn:mu_prime}  \nonumber\\
		& = (1 - p_0) \mu + p_1 (1 - \mu), \\
\nu' & = (1 - p_0) \nu + p_1 (1 - \nu). \label{eqn:nu_prime}
\end{align}

\begin{figure}[H]
\centering
\includegraphics[width=2.3in]{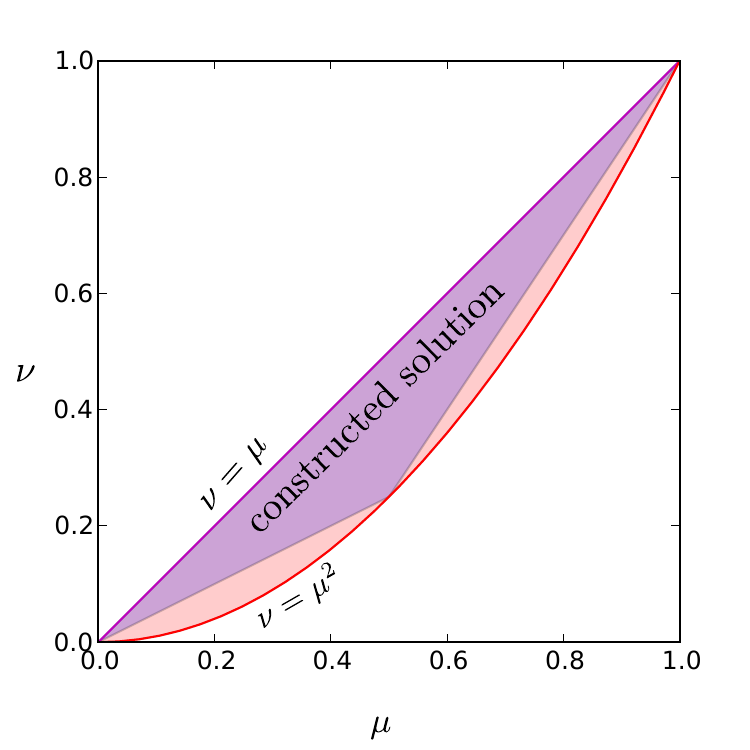}
	\caption{The full shaded region includes all allowed values for the constraints $\mu$ and $\nu$ for all possible probability distributions, replotted from Figure~\ref{fig:allowed}.  As described in Appendices~\ref{sec:prime_construction} and \ref{sec:validity}, one of our low-entropy constructed solutions can be matched to any of the allowed constraint values in the full shaded region, whereas the constructed solution described in Appendix~\ref{sec:comm_construction} can achieve any of the values within the triangular purple shaded region.
Note that even with this second solution, we can cover most of the allowed region.
Each of our constructed solutions have entropies that scale as $S \sim \ln(N)$.}
\label{fig:constructed}
\end{figure}

These values span a~triangular region in the $\mu$-$\nu$ plane that covers the majority of satisfiable constraints. Figure~\ref{fig:constructed} illustrates the situation for $\mu = \nicefrac{1}{2}$. Note that by starting with other values of $\mu$, we can construct a~low entropy solution for any possible constraints $\mu'$ and $\nu'$.

With the addition of these two states, the entropy of the expanded system $\tilde{S}^{con2'}_2$ is bounded from above by
\begin{equation}
\tilde{S}^{con2'}_2 = p_n \tilde{S}^{con2}_2 - \sum_{i \in \{0,1,n\}} p_i \log_2 (p_i) \label{eqn:ext_Scon2}
\end{equation}

For given values of $\mu'$ and $\nu'$, the $p_i$ are fixed and only the first term depends on $N$.
Using~Equations~\eqref{eqn:mu_prime} and \eqref{eqn:nu_prime},
\begin{equation}
p_n = \frac{\mu' - \nu'}{\mu - \nu}.
\end{equation}

This allows us to rewrite Equation~\eqref{eqn:ext_Scon2} as
\begin{equation}
\tilde{S}^{con2'}_2 \leq  \left( \frac{\mu' - \nu'}{\mu - \nu} \right) \tilde{S}^{con2}_2 + \log_2 (3).
\end{equation}

We are free to select $\mu$ and $\nu$ to minimize the first coefficient for a~desired $\mu'$ and $\nu'$, but in general we know this coefficient is less than 1 giving us a~simple bound,
\begin{equation}
\tilde{S}^{con2'}_2 \leq   \tilde{S}^{con2}_2 + \log_2 (3).
\end{equation}

Like the original distribution, the entropy of this distribution scales logarithmically with $N$. Therefore, by picking our original distribution properly, we can find low entropy distributions for any $\mu$ and $\nu$ for which the number of active states grows as a~polynomial in $N$ (see Figure~\ref{fig:constructed}).

Similarly, we can extend the range of validity for the construction described in Appendix~\ref{sec:comm_construction}  to the triangular region shown in Figure~\ref{fig:allowed} by assigning probabilities $p_0$, $p_1$, and $p_{N/2}$ to the all silent state, all active state, and the total probability assigned to the remaining $2N - 2$ states of the original model, respectively. The entropy of this extended distribution must be no greater than the entropy of the original distribution (Equation~\eqref{eq:S_con}), since the same number of states are active, but now they are not weighted equally, so this remains a~low entropy distribution.

\section{Proof of the Lower Bound on Entropy for Any Distribution Consistent with Given \boldmath{$\{\mu_i\}$} \& \boldmath{$\{\nu_{ij}\}$}}
\label{sec:lower_bound}

Using the concavity of the logarithm function, we can derive a~{\em lower} bound on the minimum entropy.  Our lower bound asymptotes to a~constant except for the special case $\mu_i = \nicefrac{1}{2}$,  $\forall$ $i$, and~\mbox{$\nu_{ij} = \nicefrac{1}{4}$}, $\forall$ $i,j$, which is especially relevant for communication systems since it matches the low order statistics of the global maximum entropy distribution for an unconstrained set of binary~variables.

We begin by bounding the entropy from below as follows:
\begin{align}
S({\mb p})	& = - \sum_{w} p(w) \log_2 p(w)  \nonumber\\
			& = \left\langle - \log_2 p(w) \right\rangle  \nonumber\\
			& \geq - \log_2 \left\langle p(w) \right\rangle  \nonumber\\
            & \geq - \log_2 \|{\mathbf{p}}\|_2^2  \label{eqn:entropy_bound},
\end{align}
where ${\mb p}$ represents the full vector of all $2^N$ state probabilities, and we have used $\langle \cdot \rangle$ to denote an~average over the distribution $p(w)$. The third step follows from Jensen's inequality applied to the convex function $- \log(x)$.

%
%

Now we seek an upper bound on ${\mb p}^2$.
This can be obtained by starting with the matrix representation $C$ of the constraints (for now, we consider each state of the system, $\vec{s}_i$, as a~binary column vector, where $i$ denotes the state and each of the $N$ components is either 1 or 0):
\begin{align}
C	& = \avg{\vec{s} \vec{s}^T}  \nonumber\\
	& = \sum_i p(s_i) \vec{s}_i \vec{s}^T_i \label{eqn:def_C},
\end{align}
where $C$ is an $N\times N$ matrix.
In this form, the diagonal entries of $C$, $c_{mm}$, are equal to $\mu_m$ and the off diagonal entries, $c_{mn}$, are equal to $\nu_{mn}$.

For the calculation that follows, it is expedient to represent words of the system as $\vec{\bar{s}} \in \{-1, 1\}^N$ rather than $\vec{s} \in \{0, 1\}^N$ ({ i.e.}, -1 represents a~silent neuron instead of 0).
The relationship between the two can be written
\begin{equation}
\vec{\bar{s}} = 2\vec{s} - \vec{1},
\end{equation}
where $\vec{1}$ is the vector of all ones.
Using this expression, we can relate $\bar{C}$ to $C$:
\begin{align}
\bar{C}		& = \avg{\vec\bar{s} \vec{\bar{s}}^T}  \nonumber\\
			& = \avg{(2\vec{s} - \vec{1})(2\vec{s}^T - \vec{1}^T)} \nonumber \\
			& = 4 \avg{\vec{s}\vec{s}^T} - 2 \avg{\vec{s} \vec{1}^T} - 2 \avg{\vec{1} \vec{s}^T} + \vec{1}\vec{1}^T, \\
\bar{c}_{mn}& = 4 c_{mn} - 2 c_{mm} -2 c_{nn}  + 1.
\end{align}

This reduces to
\begin{align}
\bar{c}_{mm}	& = 1, \label{eq:cbarmmresult}\\
 \bar{c}_{mn}	& = 4\nu_{mn} - 2(\mu_m + \mu_n) + 1, \quad  m \neq n  .
 \label{eq:cbarmnresult}
\end{align}

Returning to Equation~\eqref{eqn:def_C} to find an upper bound on ${\mb p}^2$, we take the square of the Frobenius norm of $\bar{C}$:
\begin{align}
\|\bar{C}\|^2_F	& = \operatorname{Tr}(\bar{C}^T\bar{C}) \label{eq:Frbenius}\nonumber\\
\begin{split}
				& = \operatorname{Tr} \left( \left(\sum_i p(\vec{\bar{s}}_i) \vec{\bar{s}}_i \vec{\bar{s}}^T_i \right) \right. \times \left. \left(\sum_j p(\vec{\bar{s}}_j) \vec{\bar{s}}_j \vec{\bar{s}}^T_j\right) \right)
\end{split}  \nonumber\\
				& = \operatorname{Tr}\left( \sum_{i,j} p(\vec{\bar{s}}_i) p(\vec{\bar{s}}_j) \vec{\bar{s}}_i \vec{\bar{s}}^T_i \vec{\bar{s}}_j \vec{\bar{s}}^T_j \right)  \nonumber\\
				& = \sum_{i,j} p(\vec{\bar{s}}_i) p(\vec{\bar{s}}_j) \operatorname{Tr}\left( \vec{\bar{s}}_i \vec{\bar{s}}^T_i \vec{\bar{s}}_j \vec{\bar{s}}^T_j \right)  \nonumber\\
        & = \sum_{i,j} p(\vec{\bar{s}}_i) p(\vec{\bar{s}}_j) \operatorname{Tr}\left( \vec{\bar{s}}^T_j \vec{\bar{s}}_i \vec{\bar{s}}^T_i \vec{\bar{s}}_j  \right)  \nonumber\\
        & = \sum_{i,j} p(\vec{\bar{s}}_i) p(\vec{\bar{s}}_j) \left( \vec{\bar{s}}_i \cdot \vec{\bar{s}}_j \right)^2  \nonumber\\
				& \geq \sum_i p(\vec{\bar{s}}_i)^2 \left( \vec{\bar{s}}_i \cdot \vec{\bar{s}}_i \right)^2  \nonumber\\
				& = N^2 \|{\mathbf{p}}\|_2^2.
\end{align}

The final line is where our new representation pays off: in this representation, $\vec{\bar{s}}_i \cdot \vec{\bar{s}}_i = N$.
This~gives us the desired upper bound for ${\mb p}^2$:
\begin{equation}
\frac{\| \bar{C} \|_F^2}{N^2} \geq \|{\mathbf{p}}\|_2^2 \label{eqn:p_bound}.
\end{equation}

Using Equations~\eqref{eq:cbarmmresult}--\eqref{eq:Frbenius}, we can express $\|\bar{C}\|_F^2$ in terms of $\mu$ and $\nu$:
\begin{align}
\|\bar{C}\|^2_F	& = \sum_m \bar{c}^2_{mm} + \sum_{m \neq n} \bar{c}^2_{mn} \nonumber \\
				& = N + (4\nu_{mn} - 2(\mu_m + \mu_n) + 1)^2 .
\end{align}

Combining this result with Equations~\eqref{eqn:p_bound} and \eqref{eqn:entropy_bound}, we obtain a~lower bound for the entropy for any distribution consistent with any given sets of values $\{ \mu_i\}$ and $\{ \nu_{ij}\}$:
\begin{align}
S({\mb p}) \geq \tilde{S}^{lo}_2 & =  \log_2 \left(\frac{N^2}{N + \sum_{i \neq j} \alpha_{ij}} \right)  \nonumber\\
& = \log_2 \left(\frac{N}{1 + (N-1) \bar\alpha} \right)
\end{align}
where $\alpha_{ij} = (4\nu_{ij} - 2(\mu_i + \mu_j) + 1)^2$ and $\bar\alpha$ is the average value of $\alpha_{ij}$ over all $i,j$ with $i \neq j$.

In the case of uniform constraints, this becomes
\begin{equation}
S({\mb p}) \geq \tilde{S}^{lo}_2 =  \log_2 \left(\frac{N}{1 + (N-1) \alpha } \right),
\end{equation}
where $\alpha = (4(\nu - \mu) + 1)^2$.

For large values of $N$ this lower bound asymptotes to a~constant
\begin{equation}
\lim_{N \rightarrow \infty}\tilde{S}_2^{lo} = \log_2 \left(1/ \bar \alpha \right).
\end{equation}

The one exception is when \(\bar \alpha = 0\).
In the large $N$ limit, this case is limited to when \(\mu_i = \nicefrac{1}{2}\) and \(\nu_{ij} = \nicefrac{1}{4}\) for all \(i\), \(j\).
Each \(\alpha_{ij}\) is positive semi-definite; therefore, \(\bar \alpha = 0\) only when each \(\alpha_{ij} = 0\).
\mbox{In other words,}
\begin{equation}
4 \nu_{ij} - 2( \mu_i + \mu_j ) + 1 = 0 \quad \forall i \neq j
\end{equation}

But in the large $N$ limit,
\begin{equation}
\mu_i \mu_j \leq \nu_{ij} \leq \min(\mu_i, \mu_j).
\end{equation}

Without loss of generality, we assume that \(\mu_i \leq \mu_j\).
In this case,
\begin{align}
0 & \leq  \mu_i \leq \frac{1}{2} \\
\frac{1}{2} & \leq \mu_j \leq \mu_i + \frac{1}{2},
\end{align}
and
\begin{equation}
\nu_{ij} = \frac{\mu_i + \mu_j}{2} - \frac{1}{4}. \label{eqn:nu}
\end{equation}

Of course, that means that in order to satisfy \(\bar \alpha = 0\) each pair must have one \(\mu\) less than of equal to \(\nicefrac{1}{2}\) and the other greater than or equal to \(\nicefrac{1}{2}\).
The only way this may be true for all possible pairs is if all \(\mu_i\) are equal to \(\nicefrac{1}{2}\).
According to Equation~\eqref{eqn:nu}, all \(\nu_{ij}\) must then be equal to \(\nicefrac{1}{4}\).
This is precisely the communication regime, and in this case our lower-bound scales logarithmically with \(N\),
\begin{equation}
\tilde{S}^{lo}_2 = \log_2 (N).
\end{equation}





\reftitle{References}


\end{document}